\documentclass[aps,prd,showpacs,preprintnumbers,amsmath,amssymb,superscriptaddress,nofootinbib,11pt]{revtex4}
\usepackage{dcolumn}
\usepackage[colorlinks,dvipdfm]{hyperref}
\usepackage[dvips]{graphicx}
\usepackage{hyperref,xcolor}
\definecolor{wine-stain}{rgb}{0.5,0,0}
\hypersetup{
  colorlinks,
  linkcolor=wine-stain,
  linktoc=all
}

\newcommand{\nn}{\nonumber}
\newcommand{\be}{\begin{equation}}
\newcommand{\ee}{\end{equation}}
\newcommand{\ba}{\begin{array}}
\newcommand{\bqa}{\begin{eqnarray}}
\newcommand{\eqa}{\end{eqnarray}}
\newcommand{\cO}{{\cal O}}
\newcommand{\mL}{\mathcal{L}}

\newcommand{\ket}{\,\rangle}
\newcommand{\bra}{\langle \,}

\newcommand{\wwL}{\widetilde{\widetilde{L}}}

\newcommand\lsim{\mathrel{\rlap{\lower4pt\hbox{\hskip1pt$\sim$}}
    \raise1pt\hbox{$<$}}}
\newcommand\gsim{\mathrel{\rlap{\lower4pt\hbox{\hskip1pt$\sim$}}
    \raise1pt\hbox{$>$}}}
\newcommand{\ea}{\end{array}}
\newcommand{\bear}{\begin{small}\begin{eqnarray}}
\newcommand{\eear}{\end{eqnarray}\end{small}}
\def\bat{\begin{array}{cc}}

\newcommand{\Frac}[2]{\frac{\displaystyle #1}{\displaystyle #2}}
\newcommand{\Int}{\displaystyle{\int}}
\def\ie{{\it i.e.},\ }

\begin{document}

\vspace*{-0.6cm}
\hspace*{10.5cm}{\small   {FTUAM-14-4}  }
\\
\vspace*{-0.6cm}
\hspace*{11cm}{\small {IFT-UAM/CSIC-14-010} }
\\
\vspace*{2cm}

\title{   Resonance effects   in pion and kaon decay constants
}

\author{Zhi-Hui~Guo \footnote{zhguo@mail.hebtu.edu.cn}   }

\affiliation{ Department of Physics, Hebei Normal University, 050024 Shijiazhuang, P.R.China  \\
 and State Key Laboratory of Theoretical Physics, Institute of Theoretical Physics, Chinese Academy of Sciences, Beijing 100190, P.R.China }

\author{ Juan~Jos\'e~Sanz-Cillero  \footnote{juanj.sanz@uam.es}   }

\affiliation{  Departamento de F\'\i sica Te\'orica and Instituto de F\'\i sica Te\'orica, IFT-UAM/CSIC, Universidad Aut\'onoma de Madrid, Cantoblanco, 28049 Madrid, Spain }

\begin{abstract}
 In this article we study impact of the lightest vector and scalar resonance multiplets
in the pion and kaon decay constants up to next-to-leading order in the $1/N_C$ expansion, \ie
up to the one-loop level.
The $F_\pi$ and $F_K$ predictions  obtained within the framework of Resonance Chiral Theory
are confronted with lattice simulation data.
The vector loops (and  the $\rho-\pi\pi$ coupling $G_V$ in particular)
are found to play a crucial role in the
determination of the Chiral Perturbation Theory couplings  $L_4$ and $L_5$ at next-to-leading
order in $1/N_C$. Puzzling,  values of $G_V\lsim 40$~MeV seem to be necessary to agree
with current  phenomenological results for $L_4$ and $L_5$. Conversely, a value of
$G_V\gsim 60$~MeV compatible with standard $\rho-\pi\pi$ determinations
turns these chiral couplings negative. However, in spite of the strong anti-correlation with $L_4$,
the $SU(3)$ chiral coupling $F_0$ remains stable all the time and
stays within the range $78 \sim 86$~MeV when  $G_V$ is varied in a wide range, from $40$ up to $70$~MeV.
%
%
Finally, we would like to remark that the leading order expressions used in this article for
the $\eta-\eta'$ mixing, mass splitting of the vector multiplet  masses
and the quark mass dependence of the $\rho(770)$ mass
are  found in reasonable agreement with the lattice data.
\end{abstract}

\vskip .5cm


\date{\today}
\maketitle

\tableofcontents

\section{Introduction}

The  decay constants of the light pseudo Nambu-Goldstone bosons (pNGB)
$\pi$ and $K$ are important quantities in particle physics.
 Their precise determinations are crucial for the extraction of
the Cabibbo-Kobayashi-Maskawa  matrix elements $V_{ud}$ and $V_{us}$
and for beyond Standard Model physics searches  in the
flavour        sector~\cite{FLAG:2013,Bernard:2006}.
They are one of the fundamental parameters in chiral perturbation theory ($\chi$PT),
the effective field theory (EFT)  of  Quantum Chromodynamics (QCD)
that  describes the low-energy interactions between the pNGB ($\pi, K, \eta$) from the spontaneous
chiral symmetry breaking~\cite{Weinberg:1979,gl845}.
In fact, these two decay constants $F_\pi$ and $F_K$
have been widely studied in $\chi$PT phenomenology~\cite{gl845,Bijnens:2011tb}
and  lattice simulations~\cite{FLAG:2013,Colangelo:2010et}.
However due to the rapid proliferation of the number of unknown low energy constants
(LECs)  at $\cO(p^6)$, it is rather difficult to extract  definitive conclusion
on the values of $\cO(p^4)$ LECs  and the $\cO(p^2)$ coupling $F_0$~\cite{Bijnens:2011tb,FLAG:2013}.

In spite of important progresses in the last years, lattice simulations usually
compute the pNGB decay constants for values of the quark masses $m_q$  heavier than the physical ones,
in order to optimize  computer resources.
This worsens the convergence of the $\chi$PT series and higher chiral orders must be accounted
and resummed in an appropriate way.
However,  apart from the chiral log behaviour at small quark masses, these observables
show an almost  linear dependence on $m_q$, without any significant logarithmic behaviour  that
one would expect from hadronic loop contributions.
The inclusion of resonances within a chiral invariant framework,
Resonance Chiral Theory (R$\chi$T)~\cite{rcht89},
is expected  to extend the applicability energy region of $\chi$PT up to some higher scale
and explain this feature.
The $1/N_C$ expansion~\cite{Nc}, with $N_C$ the numbers of colors in QCD,
is taken as a guiding principle
in R$\chi$T to sort out the various contributions, being hadronic loops suppressed by $1/N_C$.
Indeed,  at leading order (LO) in $1/N_C$,
R$\chi$T predicts an almost linear $m_q$ dependence for the decay constants
with a slope given by the lightest scalar resonance
mass~\cite{SanzCillero:2004sk}, with  fit  value $M_S=1049\pm 25$~MeV:
the same scalar resonance that mediates the scalar form-factor into two pNGB at tree-level
also     rules the quark mass corrections
in the weak pNGB decay  through an axial-vector current.
%

In the  present work, we calculate the pion and kaon decay constants  up to
next-to-leading order (NLO) in $1/N_C$ within R$\chi$T, \ie  up to the one-loop level,
continuing a series of previous NLO computations in this
work-line~\cite{SanzCillero:2009ap,Rosell-L8,Rosell-L9-L10,L9a,Cata:2001nz}.
We hope in this way to properly incorporate the small $m_q$ chiral log behaviour
without spoiling the roughly linear dependence found at large $N_C$~\cite{SanzCillero:2004sk}.
This will allow us to match $SU(3)$ $\chi$PT at $\cO(p^4)$ recovering the right renormalization
scale dependence of the relevant LECs, $L_4(\mu)$ and $L_5(\mu)$.
These theoretical predictions from R$\chi$T  will be  then confronted with
the lattice results for $F_\pi$,
$F_K$~\cite{Davies:2003fw,Davies:2003ik,Aoki:2010dy,Arthur:2012opa}
and $F_K/F_\pi$~\cite{Durr:2010hr}.
%

The impact of meson resonances on the pNGB decay constants
have  not  been  thoroughly discussed in previous literature.
The only other one-loop
attempt was carried out in the $SU(2)$ case and incorporated
only the lightest scalar~\cite{Soto:2011ap}.
In this work we discuss the $SU(3)$ chiral dynamics  and the effect of vector loops, in addition to the scalar ones.
The outcomes in the present article are not expected to provide
an improved version of the already very precise $\chi$PT computations present in the market,
which are known now up to next-to-next-to-leading order (NNLO) in the chiral expansion~\cite{FP-Op6,Ecker:2013pba}
and incorporate
specific lattice simulation subtleties (twisted boundary conditions~\cite{lattice-TBC},
finite volume effects~\cite{finite-volume},   etc.).
The central aim of this article is to show how it is possible to study the dynamics   of  the lightest  resonances
through the analysis of these observables in the lattice.
In particular we will see that the vector resonance loops
  (and more precisely the $\rho-\pi\pi$ coupling $G_V$)
play an important role in the analysis and will be crucial for the final values of the
$\chi$PT LECs   $F_0$, $L_4$ and $L_5$.

The article is organized as follows: in Sec.~\ref{sec.RChT}  we introduce theoretical setup
and the LO and NLO  R$\chi$T Lagrangian.
In Sec.~\ref{sec.theory} we perform the NLO computation in R$\chi$T,
renormalization and  matching between R$\chi$T and $\chi$PT.
The fit to lattice data and the phenomenological discussions are
carried out in Sec.~\ref{sec.pheno}. We finally provide the conclusions  in Sec.~\ref{sec.conclusions},
relegating the most technical details to the Appendices.

\section{Relevant R$\chi$T Lagrangian}
\label{sec.RChT}

\subsection{R$\chi$T building blocks}

We  will use   the  exponential realization of the
$U(3)_L\otimes U(3)_R/U(3)_V$ coset coordinates for the~pNGB,
\begin{eqnarray}
U &=&  u^2 = e^{i\frac{ \sqrt2\phi}{ F_0}} \,,
\qquad\qquad
D_\mu U = \partial_\mu U - i r_\mu U + i U \ell_\mu \,,
\label{eq.U-DU-def}
\end{eqnarray}
where  the covariant derivative $D_\mu U$ incorporates the right and left external
sources, respectively, $r_\mu$ and $\ell_\mu$, in such a way that it transforms
in the same way as $U$ under    local  chiral transformations~\cite{gl845}:
  \bear
U\quad \longrightarrow \quad  g_R \, U\, g_L^\dagger \, ,
\qquad\qquad\qquad
u\quad \longrightarrow \quad  g_R \, u\, h^\dagger \,\,\,= \,\,\, h\, u\, g_L^\dagger \, ,
\eear
with the compensating transformation $h(\phi,g_R,g_L)$~\cite{rcht89}.
Thus, the covariant derivative in Eq.~\eqref{eq.U-DU-def} transforms
in the form $(D_\mu U)\longrightarrow g_R \, (D_\mu U)\, g_L^\dagger$.

The pNGB octet plus the singlet $\eta_1$  are given by the matrix,
\begin{equation}\label{phi1}
\phi \,\,\,=\,\, \, \sum_{a=0}^8 \phi^a \Frac{\lambda^a}{\sqrt{2}}
\, \,\, =\,\,\, \left( \begin{array}{ccc}
\frac{1}{\sqrt{2}} \pi^0+\frac{1}{\sqrt{6}}\eta_8+\frac{1}{\sqrt{3}} \eta_1 & \pi^+ & K^+ \\ \pi^- &
\frac{-1}{\sqrt{2}} \pi^0+\frac{1}{\sqrt{6}}\eta_8+\frac{1}{\sqrt{3}} \eta_1   & K^0 \\  K^- & \bar{K}^0 &
\frac{-2}{\sqrt{6}}\eta_8+\frac{1}{\sqrt{3}} \eta_1
\end{array} \right)\,.
\end{equation}
Notice that due to the inclusion of the singlet $\eta_1$, the standard
  chiral counting from $SU(3)$--$\chi$PT given by    an expansion in powers of
the momenta and the pNGB masses      
does not work any more,
since the mass of $\eta_1$ does not vanish in the chiral limit
($m_{\eta_1}\to M_0\simeq 850$~MeV when $m_q\to 0$~\cite{Feldmann:1999uf}).
However, by introducing
$1/N_C$ as a third expansion parameter,
it is still possible to establish a consistent power
counting system for $U(3)$--$\chi$PT~\cite{Kaiser:2000gs}, which includes the singlet $\eta_1$ as a dynamical degree of freedom (d.o.f).

The basic building blocks of the meson theory read
\begin{eqnarray}\label{defbb}
u_\mu = i u^\dagger  D_\mu U u^\dagger \, =\,
i \{ u^\dagger (\partial_\mu - i r_\mu) u\, -\, u(\partial_\mu - u\ell_\mu) u^\dagger\}
\,, \nn\\
\chi_\pm  = u^\dagger  \chi u^\dagger  \pm  u \chi^\dagger  u \,,
\qquad \qquad f_\pm^{\mu\nu} = u F_L^{\mu\nu} u^\dagger \, \pm \,
u^\dagger F_R^{\mu\nu} u\, ,
\end{eqnarray}
where $\chi=2 B (s + i p)$  includes the scalar ($s$) and pseudo-scalar ($p$)
external sources, and  $F_{L}^{\mu\nu}$ and $F_{R}^{\mu\nu}$  are, respectively,
the left and right    field-strength   tensors~\cite{gl845}.
All the referred tensors $X=u_\mu,\, \chi_\pm,\, f_\pm^{\mu\nu}$ transform
   under chiral transformations as
\be
X\quad \longrightarrow \quad  h\, X\, h^\dagger\, .
\label{eq.X-transformation}
\ee
We will also make use of the covariant derivative    for this type of objects,
\begin{eqnarray}
\nabla_\mu X &=& \partial_\mu X + [\Gamma_\mu, X]\,,
\qquad\qquad \Gamma_\mu  = \frac{1}{2}\bigg[ u^\dagger (\partial_\mu- i\,r_\mu) u
+ u (\partial_\mu- i\,\ell_\mu) u^\dagger \bigg]\,.
\end{eqnarray}

In our analysis we will study the impact of the lightest  $U(3)$ nonets
of  vector and scalar resonances surviving at large $N_C$. We will employ a representation of the resonance
fields $R=V,\, S$
such that they transform in the way $ R\longrightarrow h \, R\, h^\dagger$ in Eq.~\eqref{eq.X-transformation}
under chiral
transformations~\cite{rcht89}.
The flavor assignment for the scalar and vector resonances is similar
to that in Eq.~\eqref{phi1}:
\begin{eqnarray}
\label{s1s8}
S &=& \left(
                                        \begin{array}{ccc}
                                          \frac{a_0^0}{\sqrt2}+\frac{\sigma_8}{\sqrt6}+\frac{\sigma_1}{\sqrt3} & a_0^+ & \kappa^{+}  \\
                                          a_0^- & -\frac{a_0^0}{\sqrt2}+\frac{\sigma_8}{\sqrt6}+\frac{\sigma_1}{\sqrt3}  & \kappa^{0}  \\
                                          \kappa^{-} & \bar{\kappa}^{0} & -\frac{2\sigma_8}{\sqrt6}+\frac{\sigma_1}{\sqrt3} \\
                                        \end{array}
                                       \right) \,, \\  \nn\\
V_{\mu\nu}&=& \left(
                                        \begin{array}{ccc}
                              \frac{\rho_0}{\sqrt2}+\frac{1}{\sqrt{6}}\omega_8 + \frac{1}{\sqrt{3}} \omega_1 & \rho^+ & K^{*+}  \\
                         \rho^- & -\frac{\rho_0}{\sqrt2}+\frac{1}{\sqrt{6}}\omega_8+ \frac{1}{\sqrt{3}}\omega_1   & K^{*0}  \\
                K^{*-} & \bar{K}^{*0} & -\frac{2}{\sqrt{6}}\omega_8+ \frac{1}{\sqrt3}\omega_1 \\
                                        \end{array}
                                       \right)_{\mu\nu} \,.
\end{eqnarray}
The vector resonances are described here in the antisymmetric
tensor formalism through the $V_{\mu\nu}$ fields~\cite{rcht89}.
In later discussions, we will consider the ideal $I=0$ resonance  mixings
\begin{eqnarray}
\sigma_8=\sqrt{\frac{1}{3}}\sigma-\sqrt{\frac{2}{3}}\sigma'\,, \quad
\sigma_1=\sqrt{\frac{2}{3}}\sigma+\sqrt{\frac{1}{3}}\sigma'\,,
\end{eqnarray}
\begin{eqnarray}
\omega_8 = \sqrt{\frac{2}{3}} \phi + \sqrt{\frac{1}{3}} \omega \,, \quad
\omega_1 = \sqrt{\frac{2}{3}} \omega - \sqrt{\frac{1}{3}} \phi \,,
\end{eqnarray}
for the octet and singlet scalar and vector resonances,
which leads to two different types of isoscalar resonances
$R^{\bar{u}u+\bar{d}d}_{I=0}$ and  $R^{\bar{s}s}_{I=0}$ in the quark flavour basis.
This pattern was found to provide an excellent phenomenological description for the
    vector    resonance  multiplets~\cite{Guo:2009hi}.
We would like to stress that the resonances incorporated
in our framework are the ones surviving at large $N_C$.
 The   lowest multiplet of vector resonances
($\rho, K^*, \omega, \phi$)   behaves very approximately like a  standard
$\bar{q}q$ resonance, with a mass that tends to a constant and a width decreasing
like  $1/N_C$ when
$N_C\to \infty$~\cite{RuizdeElvira:2010cs,Guo:2011pa,Guo:2012ym,Guo:2012yt}.
This allows us to build a one-to-one correspondence between the physical vector resonances and those surviving at large $N_C$.
On the other hand,  the nature of the light scalar resonances,
such as $f_0(500)$,$f_0(980)$, $K^*_0(800)$, etc.,
is still unclear and various descriptions  are proposed by different groups:
meson-meson molecular, tetraquark, standard $\bar{q}q$ with a strong pion cloud, etc.
As a result of this, their $N_C$ behavior is also
under debate~\cite{Guo:2011pa,Guo:2012yt,RuizdeElvira:2010cs,Dai:2011bs,Dai:2012kf,Zhou:2010ra}.
Though the $N_C$ trajectories of the scalar resonances reported by different groups
diverge from each other,
surprisingly there is one common feature from Refs.~\cite{Guo:2011pa,Guo:2012yt,RuizdeElvira:2010cs,Dai:2011bs,Dai:2012kf}:
 a scalar resonance with mass around 1~GeV appears  at large $N_C$.
Based on these results and the success of this hypothesis in previous
analyses~\cite{Guo:2009hi,Rosell-L8,SanzCillero:2009ap},
we will assume in the present article the existence
of a large--$N_C$ scalar nonet with a bare mass around 1~GeV.

On the other hand, the situation is slightly more cumbersome
for $\eta_8$ and $\eta_1$ and one needs to consider the mixing
\begin{eqnarray}\label{etamixing}
\eta_8=c_\theta\eta+s_\theta\eta'\,,  \quad
\eta_1=-s_\theta\eta+c_\theta\eta'\,,
\end{eqnarray}
with $c_\theta=\cos{\theta}$ and $s_\theta=\sin{\theta}$.
Phenomenologically, one has
$\theta= (-13.3 \pm  0.5)^\circ$ in QCD~\cite{theta-exp},
far away from the ideal mixing $\theta= -\arcsin{\sqrt{\frac{2}{3}}} \simeq
-55^\circ$.
We will see that only the leading order mixing
will be relevant in the present analysis of $F_\pi$
and $F_K$.~\footnote{
This is because $\eta$ and $\eta'$ only enter the pion
and kaon decay constants through the chiral loops.
Subleading contributions to the mixing will be neglected
as they will enter as corrections  in  one-loop suppressed
diagrams in the pNGB decay.       }
In the loop calculation, it is convenient to use the physical states
$\eta$ and $\eta'$,
instead of the flavour eigenstates $\eta_1$ and $\eta_8$.
The reason is that the mixing between $\eta_1$ and $\eta_8$
is proportional to
$m_K^2-m_\pi^2$, which is formally the same order as the masses
of $\eta_1$ and $\eta_8$. The insertion of the $\eta_1$ and $\eta_8$
mixing in the chiral loops will not increase the $1/N_C$
order of the loop diagrams.    This  makes the loop calculation
technically  complicated. However, as already noticed in
Refs.~\cite{Guo:2011pa,Guo:2012ym,Guo:2012yt},
 one can easily avoid the complication in the loop computation
by  expressing the Lagrangian in terms of the $\eta$ and $\eta'$ states
resulting from the diagonalization of $\eta_1$ and $\eta_8$ at leading order.
In addition, the effect of the mixing is less and less important
in the lattice simulations as
$m_\pi$ increases and approaches   $m_K$, making subleading uncertainties
in the mixing even more suppressed.
Therefore, in the following
discussion, we will always calculate the loop diagrams
in terms of   $\eta$ and $\eta'$ states, instead of $\eta_1$ and $\eta_8$.
Further details on  the $\eta$--$\eta'$ mixing are relegated to
App.~\ref{app.eta-etap}.

\subsection{ LO  Lagrangian}

In general, one can classify the R$\chi$T operators in the
Lagrangian according to the number of resonance fields  in the form
\bear
\mL_{R\chi T} &=& \mL_G\quad +\quad \sum_R \mL_R\quad +\quad ...
\eear
where the operators in $\mL_G$ only contains pNGB and external sources,
the $\mL_R$ terms have one  resonance field  in addition to possible pNGB
and external auxiliary fields,  and the dots stand for operators with
two or more resonances.

 We focus first on the $\mL_G$ part of the R$\chi$T Lagrangian.
Since we will later incorporate the lightest $U(3)$ nonet of  hadronic resonances
and we are working within a large--$N_C$ framework,
our theory will be based on the $U(3)_L\otimes U(3)_R$ symmetry
and, in addition to the two usual $\cO(p^2)$ operators from
$SU(3)$ $\chi$PT,  we will also need to consider the singlet $\eta_1$
mass term:
\begin{eqnarray} \label{lolagrangian}
  \mL_{G}^{\rm LO}
\,\,\, =\,\,\, \frac{\widetilde{F}^2}{4}\langle u_\mu u^\mu \rangle \,
+\frac{\hat{F}^2}{4}\langle \chi_+ \rangle
+ \frac{F_0^2}{3}M_0^2 \ln^2{\det u}\,,
\end{eqnarray}
where  $\langle \ldots \rangle$
stands for the trace in flavor space.
The last operator in the right-hand side (r.h.s.) of
Eq.~\eqref{lolagrangian} is generated
by the $U_A(1)$ anomaly and gives mass to the singlet $\eta_1$.
On the contrary to $\chi$PT,
in R$\chi$T one generates ultraviolet (UV) divergences
which require the first two terms in the r.h.s
of Eq.~(\ref{lolagrangian}) to fulfill the renormalization of the resonance
loops~\cite{SanzCillero:2009ap,RChT-gen-fun}.
Notice that a different coupling notation $\alpha_1=\widetilde{F}^2/4$ and
$\alpha_2= \hat{F}^2/4$ is used in Ref.~\cite{RChT-gen-fun}.
    As $\widetilde{F}$ and $\hat{F}$  describe the chiral limit pNGB decay
constant from an axial-vector current and a pseudo-scalar density, respectively,
one has that $\lim_{N_C\to \infty}  \widetilde{F}/F_0=\lim_{N_C\to \infty} \hat{F}/F_0=1$.
$F_0$ stands for the $n_f=3$ decay constant of the pNGB octet
in the chiral limit. The parameter
$B$   in $\chi_+$ from Eq.~\eqref{defbb}
is connected with the quark condensate through
$\langle 0|\bar{q}^iq^j|0\rangle =- F_0^2   B     \delta^{ij}$
in the same limit.
 The explicit   chiral symmetry      breaking  is realized by setting
the scalar external source field to  $s = {\rm Diag}(m_u,m_d,m_s)$,
being $m_q$ the light quark masses. We will consider the isospin limit
all along the work, \ie
we will take $m_u = m_d$ (denoted just as $m_{u/d}$) and neglect any electromagnetic correction.

In order to account for the resonance effects, we consider the minimal resonance
operators  in the leading order   R$\chi$T Lagrangian~\cite{rcht89}
\begin{eqnarray}
\label{lagvector}
\mL_{V}     &=&  \frac{F_V}{2\sqrt2}\langle V_{\mu\nu} f_+^{\mu\nu}\rangle +
\frac{i G_V}{2\sqrt2}\langle V_{\mu\nu}[ u^\mu, u^\nu]\rangle \,,
\\
\label{lagscalar}
 \mL_{S}    &=& c_d\bra S u_\mu u^\mu \ket + c_m \bra S \chi_+ \ket\,.
\end{eqnarray}
In general, one could consider the resonance operators of the type $\bra R\chi^{(n\ge4)}(\phi) \ket$, with
the chiral tensor $\chi^{{(n)}}(\phi)$ only including the pNGB and external fields and $n$ standing for the chiral order of this
chiral tensor. The resonance operators in the previous two equations
are of  type $\bra R\chi^{(2)}(\phi) \ket$.
Operators with higher values   of $n$
 tend to  violate   the  high-energy asymptotic behaviour
dictated by QCD  for form-factors and Green functions.
%
Likewise, by means of meson field redefinitions it is possible to trade some
resonance  operators by other terms with a lower number of derivatives and
operators without resonance fields~\cite{RChT-RGE,SanzCillero:2009ap,L9a,EoM}.
As a result of this,  only the lowest order chiral tensors        
are  typically employed to build the operators of the leading order R$\chi$T Lagrangian.
We will follow  this heuristic  rule in the present work.
Nevertheless, we remind the reader that the truncation of the infinite tower of large-$N_C$ resonances introduces in general a theoretical uncertainty in the determinations, which will be neglected in our computation. Considering only the lightest resonance multiplets may lead
to some issues with the short-distance  constraints and the low-energy predictions
when a broader and broader set of observables is analyzed~\cite{truncation}.

{  In Ref.~\cite{Guo:2011pa}, two additional resonance operators
were taken into account (last two terms in Eq.~(5) of the previous reference).
These two terms are $1/N_C$ suppressed with respect to the  R$\chi$T operator in Eq.~\eqref{lagscalar}.
They happen to be irrelevant for our current study up to NLO in $1/N_C$ since
they involve at least one $\eta$ or $\eta'$ fields.
As we already mentioned previously, $\eta$ and $\eta'$ only enter our calculation through  chiral loops
and the two additional operators in Ref.~\cite{Guo:2011pa} would contribute to $F_\pi$ and $F_K$ at
next-to-next-to-leading order in $1/N_C$.
Thus, the one-loop calculation at NLO in $1/N_C$ only requires the consideration of the LO resonance operators like
those  in Eqs.~\eqref{lagvector} and~\eqref{lagscalar}.    }

The corresponding kinematical terms for resonance fields are~\cite{rcht89}
\begin{eqnarray}\label{lagkinv}
\mathcal{L}_{\rm kin}^V&=&-{1\over 2} \bra \nabla^\lambda
V_{\lambda\mu}\nabla_\nu V^{\nu\mu}-{1\over 2}\overline{M}^2_V V_{\mu\nu}V^{\mu\nu} \ket \,, \\
\mathcal{L}_{\rm kin}^S&=&{1\over 2} \bra \nabla^\mu S \nabla_\mu S
-\overline{M}^2_{S} S^{2} \ket \,.
\label{lagkins}
\end{eqnarray}

In our current work, we also incorporate
the light quark mass corrections to the resonance masses
and in the large $N_C$ limit this effect is governed
by the operators~\cite{MR-split}~\footnote{
Notice that the different canonical normalization of the scalar
and vector mass terms is the responsible of  the $(-\frac{1}{2})$ factor
in front of the vector splitting operator.     }
\begin{eqnarray}
\label{lagemr}
\mL_{RR}^{\rm split} &=&
e_m^S \bra SS \chi_+ \ket
\,\,\, -\,\,\,   \Frac{1}{2} \, e_m^V \bra V_{\mu\nu} V^{\mu\nu} \chi_+ \ket \,.
\end{eqnarray}
In the notation of Ref.~\cite{Op6-RChT} these two couplings would
be given by $e_m^S=\lambda_3^{SS}$ and $e_m^V=-2\lambda_6^{VV}$.
%
If no further bilinear resonance term is included in the Lagrangian,
one has an ideal mixing for the two $I=0$ resonances in the nonet  and
a mass splitting pattern of  the form
\begin{eqnarray}\label{masssplit}
(M^{\bar{u}u+\bar{d}d}_{I=0})^2=M^2_{I=1} &=& \overline{M}_R^{\,\,2} - 4 e_m^R m_\pi^2\,,
\nonumber\\
M^2_{I=\frac{1}{2}} &=& \overline{M}_R^{\,\,2} - 4 e_m^R m_K^2\,\,\,  \,,
\nonumber\\
M^{(\bar{s}s)\,\,\, 2}_{I=0} &=& \overline{M}_R^{\,\,2}
- 4 e_m^R \,\, (2 m_K^2-m_\pi^2)\,\,\,  \,,
\end{eqnarray}
with $\overline{M}_R$ the resonance mass in chiral limit.
Notice that in the following we will use the notations $M_S$ and $M_V$ for the
masses of scalar and vector multiplets in chiral limit, respectively.

{ At large $N_C$, the coupling of the LO Lagrangian  scale like
$F_0,\,\widetilde{F},\, \hat{F},\, G_V,\, c_d,\, c_m =\, \cO(N_C^\frac{1}{2})$ and the masses of the mesons
considered  here behave like $m_\phi,\, M_R,\, =\,\cO(N_C^0)$,
with the splitting parameter $e_m^R\, =\,\cO(N_C^0)$.  The $\eta_1$ mass
chiral limit $M_0$ is formally $\cO(N_C^{-1})$, although numerically
provides a sizable contribution to the $\eta-\eta'$ mixing
that needs to be taken into account in order to properly reproduce
their masses and mixing angles. More details can be found
in App.~\ref{app.eta-etap}.      }

\subsection{NLO  R$\chi$T  Lagrangian}

In general, one should also take into account
local operators  {  with  }   a  higher number of derivatives (e.g. $\cO(p^4)$) in R$\chi$T.
In particular one might consider  operators composed only of
pNGB and external fields.
Notice that these terms of the R$\chi$T Lagrangian
are different from those in $\chi$PT, as they are two different quantum
field theories with different particle {  content.   }

Based on phenomenological analyses and short-distance constraints
it is well known that the leading parts of the $\chi$PT LECs
are found to be saturated by the lowest resonances
at large $N_C$~\cite{rcht89,SD-RChT}.
{  The operators of R$\chi$T without resonances of $\cO(p^{d\geq4})$
can be regarded as $1/N_C$ suppressed residues, absent when $N_C\to \infty$.   }
Nonetheless the resonance saturation scale
{  cannot be determined  at large $N_C$  as this is a NLO  effect in $1/N_C$.   }
Since in this work we perform the discussion at the NLO of $1/N_C$,
{  we will include these residual R$\chi$T  operators without resonance fields,
which start being relevant at NLO in $1/N_C$.}

The pertinent $\cO(p^4)$ operators in our study are~\cite{gl845}
\begin{eqnarray} \label{lagchpt4}
   \mL_{G}^{\rm NLO}      
&=&   \widetilde{L}_4 \bra u_\mu u^\mu \ket \bra \chi_+ \ket + \widetilde{L}_5  \bra u_\mu u^\mu \chi_+ \ket
 + \widetilde{L}_6 \bra \chi_+\ket \bra \chi_+ \ket + \widetilde{L}_7\bra \chi_- \ket \bra\chi_-\ket
 \nonumber \\&& + \frac{\widetilde{L}_8}{2} \bra \chi_+\chi_+ +\chi_-\chi_-\ket
+i \,\widetilde{L}_{11}\bra \chi_-\big(\nabla_\mu u^\mu - \frac{i}{2} \chi_-
+ \frac{i}{2 {  n_f  }}\bra \chi_- \ket \big) \ket
\nonumber \\ &&
-\widetilde{L}_{12}\bra
    \big(\nabla_\mu u^\mu - \frac{i}{2} \chi_- + \frac{i}{2 {  n_f  }}\bra \chi_- \ket \big)^2
\ket\,,
\end{eqnarray}
{    where  $n_f=3$  and   the tilde is introduced to
distinguish the R$\chi$T couplings from the $\chi$PT LECs $L_j$.
The set of  $\mL_G^{\rm NLO}$ couplings scale
like $\widetilde{L}_j\,=\, \cO(N_C^0)$ within the $1/N_C$ expansion
and are suppressed with respect to the $\cO(p^4)$ LECs, which  behave like
$L_j\,=\, \cO(N_C)$.  The parameters  $\widetilde{L}_{11}$    and $\widetilde{L}_{12}$    }
will not appear
in the final results {  for   } $F_\pi$ and $F_K$, as their contributions in the
matrix element of the axial-vector current will be canceled out by the
wave function renormalization constant of the pNGB.

One should notice that the chiral operators in the previous equation are exactly
the same as in $\chi$PT,
but the coefficients can be completely different.
   In order to extract the low-energy EFT couplings one needs to integrate out
the heavy d.o.f in the R$\chi$T action.
At tree-level, {  the $\chi$PT LECs    get two kinds   } of contributions:
one comes directly from
the $\widetilde{L}_i$ operators with only pNGB and external sources;
the other, $L_i^{\rm Res}$, comes from the tree-level resonance exchanges when
$p^2\ll M_R^2$.
Hence, the relations between the couplings in R$\chi$T and those in $\chi$PT are given
by~\cite{rcht89,SD-RChT,NLO-saturation}
\begin{eqnarray}\label{lichpt}
 L_i^{\chi{\rm PT}}= L_i^{ {\rm Res}} + \widetilde{L}_i\,.
\end{eqnarray}
{}From now on,
in order to avoid any possible confusion
we will explicitly write the superscript $\chi $PT when referring the
chiral  LECs.
  The large--$N_C$ resonance contributions to the $\cO(p^4)$ LECs were computed
in Ref.~\cite{rcht89} by integrating out the resonance in the R$\chi$T generating functional,
yielding
\begin{eqnarray}
  \left. L_4^{ {\rm Res}  }\right|_{N_C\to\infty} \, = \, 0\,,  \qquad\qquad
  \left. L_5^{ {\rm Res} }\right|_{N_C\to\infty}\,  =\,
  \left.  \frac{c_d c_m}{M_S^2}\right|_{N_C\to\infty}
  \,=\, \Frac{F_0^2}{4 M_S^2}\,,
  \label{eq.LEC-saturation}
\end{eqnarray}
where in the last equality we have used the high-energy scalar form-factor
constraint~$4 c_d c_m=F_0^2$~\cite{Jamin:2001zq}.
Other couplings in Eq.~\eqref{lagchpt4} will be  irrelevant to our  final results for  the
pion and kaon decay constants.

\subsection{Scalar resonance tadpole and the field redefinition}
\label{sec.scalar-tadpole}

Before stepping {    into  the detailed calculation,   }
we point out a subtlety about the treatment of the scalar resonance operators in Eq.~\eqref{lagscalar}.
The operator with $c_m$ coupling in this equation leads to a term that
  couples  the isoscalar scalar resonances $S_8$ and $S_1$ to the vacuum.
In other words, it generates a scalar resonance tadpole proportional
to the quark masses.
Though it is not a problem to perform the calculations with such tadpole
effects, it can be rather cumbersome.
We find it is convenient to eliminate {   it    }   at the Lagrangian level.
 This   will greatly simplify the calculation when the resonances enter
the loops.
{  Nonetheless,
at tree-level  it does not make     much difference  to eliminate the tadpole  }
at the Lagrangian level~\cite{SanzCillero:2004sk}
or just to calculate { perturbatively  }
the tadpole diagrams~\cite{Guo:2011pa,Guo:2012ym,Guo:2012yt}.

In order to eliminate the scalar tadpole effects from the Lagrangian, we make the following field redefinition for the scalar resonances
\begin{equation}
\label{shiftscalar}
S=\overline{S}+\frac{c_m}{M_S^2}\chi_+\,,
\end{equation}
with $\overline{S}$   {   being the scalar resonance fields   }
after the field redefinition. By substituting Eq.~\eqref{shiftscalar}
into Eqs.\eqref{lagscalar} and \eqref{lagkins},
one has
\bear
\mL_{\rm kin}^S \, +\,  \mL_S
&=&
{1\over 2} \bra \nabla^\mu \overline{S} \nabla_\mu \overline{S}
-M^2_{S} \overline{S}^{2} \ket
\, +\, c_d \bra \overline{S} u_\mu u^\mu \ket
\,+ \Frac{c_m}{M_S^2} \bra \nabla_\mu \overline{S} \nabla^\mu \chi_+\ket
\nn\\
&& \quad
+ \, \Frac{c_d c_m}{M_S^2} \bra \chi_+ u_\mu u^\mu \ket
\, +\, \Frac{c_m^2}{2 M_S^2}\bra \chi_+\chi_+\ket
\nn\\
&& \quad
\, +\, \Frac{c_m^2}{2 M_S^4} \bra \nabla_\mu \chi_+ \nabla^\mu \chi_+\ket
\, .
\label{lagsalarshift}
\eear
The first line has the same structure as the original Lagrangian
with $S$ replaced by $\overline{S}$ but with
the corresponding tadpole operator $c_m\bra \overline{S}\chi_+\ket$ absent.
     Instead, it has been traded out by the  derivative term
$\frac{c_m}{M_S^2} \bra \nabla_\mu \overline{S} \nabla^\mu \chi_+\ket$
at the price of the extra  operators in the second and third lines.
  We want to note  that the last operator in Eq.\eqref{lagsalarshift}
is not considered in the following discussion,
as it corresponds to the set of  local $\cO{(p^6)}$ operators
without resonance fields and contributes to the decay constants at the order of $m_q^2$,
which   {   was   }      neglected and discarded in previous section.
This kind of contributions escapes the control of the present analysis,
as there are many other types of resonance operators
(e.g. the previously mentioned $\bra S\chi^{(4)}(\phi) \ket$ type) which
would generate similar terms without resonances
after the scalar field redefinition in Eq.~\eqref{shiftscalar}
but were neglected here.
The same applies to the resonance mass splitting Lagrangian $\mL_{RR}^{\rm split}$
in Eq.~\eqref{lagemr}:  the scalar field redefinition in Eq.~\eqref{shiftscalar}
generates extra splitting operators of order $m_q^2$ and $m_q^3$
which will be neglected in this work.

  At NLO in $1/N_C$, the LO Lagrangian~\eqref{lagscalar}
induces a scalar resonance tadpole proportional to   $m_q^2$
through a  pNGB loop. In order to remove it one should perform
{another}    scalar field shift
similar to Eq.~\eqref{shiftscalar} but of the form
$\Delta S\sim \frac{c_j m_\phi^4}{16\pi^2 F_0^2 M_S^2}$, with $c_j=c_d,c_m$.
This yields a contribution to the pNGB decay constants doubly suppressed,
by $m_\phi^4$ and $1/N_C$. Hence, following the  previous considerations,
we will neglect the one-loop tadpole effects.

Hence, after performing the shift in the scalar field worked out in this section,
R$\chi$T contains operators without resonances in Eq.~\eqref{lagsalarshift} with the same structure as the
$L_j$ ones in the $\cO(p^4)$ $\chi$PT Lagrangian~\cite{gl845}.
{    Combining   }   Eqs.~\eqref{lagchpt4} and \eqref{lagsalarshift},
we have the {\it effective} couplings in R$\chi$T
\bear
\label{eq.tilde-tilde-LECs}
\widetilde{\widetilde{L_4}} \,=\,   \widetilde{L_4}\,,
\qquad\qquad
\widetilde{\widetilde{L_5}} \,=\,   \Frac{c_d c_m}{M_S^2}+ \widetilde{L_5}\,,
\qquad\qquad
\widetilde{\widetilde{L_8}} \,=\,   \Frac{c_m^2}{2 M_S^2}+ \widetilde{L_8}  \,,
\eear
{  with the couplings of the remaining $\cO(p^4)$ operators    
without resonances }
just given by $\wwL_j=\widetilde{L}_j$.
In general the double-tilde notation will refer to the coupling of the
Lagrangian operator after performing the scalar field shift
in Eq.~\eqref{shiftscalar}.
It is easy to observe that at large $N_C$ one recovers the
R$\chi$T results in Eq.~\eqref{eq.LEC-saturation}:
$\wwL_4$ and $\wwL_5$ become equal to   {the}     $\cO(p^4)$ LECs
$L_4^{\chi PT}$ and  $L_5^{\chi PT}$, respectively, as there is no other possible $\cO(p^4)$
resonance contribution of this kind after performing the $S$--shift in Eq.~\eqref{shiftscalar}.
Though $L_8$ will not enter the discussion in the pion and kaon decay constants,
for completeness we comment   that our result in Eq.~\eqref{eq.tilde-tilde-LECs}
 is consistent with the scalar contributions
in Ref.~\cite{rcht89}, as $\wwL_8$ would become equal to $L_8^{\chi PT}$ in the
large--$N_C$ limit.

\section{Theoretical calculation  }
\label{sec.theory}

\subsection{Decay constants in R$\chi$T at NLO in $1/N_C$}

\begin{figure}[!t]
\begin{center}
\includegraphics[angle=0, width=0.5\textwidth]{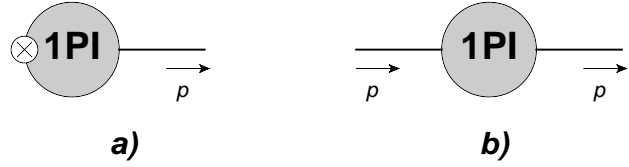}
\caption{{\small
 Relevant vertex functions for the physical pNGB decay constant:
{\bf a)}   1PI transitions between an axial-vector current
and a bare pseudo-Goldstone field, determining $F_\phi^{\rm 1PI}$;
{\bf b)} pNGB self-energy
$-i\Sigma_\phi(p^2)$.
The solid line stands for a pNGB $\phi$, the crossed circle for
an axial-vector current insertion and the circle represents all possible
1PI topologies.   }}
\label{fig.1PI-vertices}
\end{center}
\end{figure}

\begin{figure}[!t]
\begin{center}
\includegraphics[angle=0, width=0.45\textwidth]{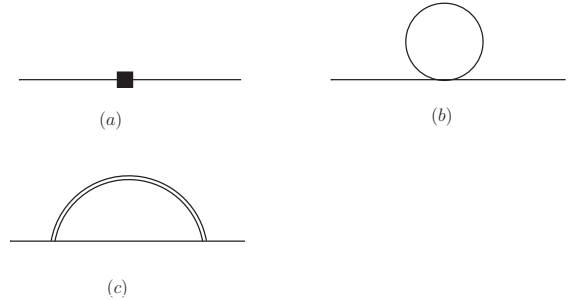}
\caption{{\small Feynman diagrams of the pNGB self energy. The single line corresponds to pNGB and the double-line
stands for {   a    }   resonance state. The tree level amplitude in diagram (a) can receive contributions both from the leading order Lagrangian in Eq.\eqref{lolagrangian}
, the contact terms appearing   {    in    }   the second line of Eq.\eqref{lagsalarshift} and  the $\cO(p^4)$ Lagrangian in Eq.\eqref{lagchpt4}.
\label{fig.se} }}
\end{center}
\end{figure}

\begin{figure}[!t]
\begin{center}
\includegraphics[angle=0, width=0.40\textwidth]{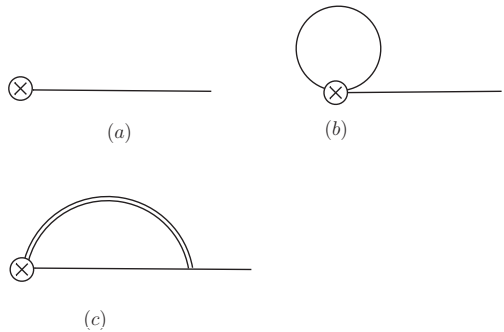}
\caption{{\small  Feynman diagrams of the pNGB axial-vector decay constant. The circled cross symbol corresponds to the axial-vector source.
Similar to Fig.~\ref{fig.se}, diagram (a) can receive contributions from the leading order Lagrangian in Eq.\eqref{lolagrangian},
the contact terms appearing    {    in    }    the second line of Eq.\eqref{lagsalarshift} and  the $\cO(p^4)$ Lagrangian in Eq.\eqref{lagchpt4}.
\label{fig.fphi} }}
\end{center}
\end{figure}

The pNGB decay constant is defined through the matrix element of the axial-vector current of the light quarks
\begin{equation}\label{deffphi}
 \bra \phi(p) | \bar{q}  \gamma_\mu \gamma_5 u | 0 \ket = - i \sqrt2 F_\phi p_\mu\,,
\qquad \qquad \mbox{with }\phi=\pi \,\, (K)\quad \mbox{for } q= d\, \,  (s)\,.
\end{equation}
In order to study the pion and kaon axial decay constants at NLO of $1/N_C$, we need to calculate the one-loop diagrams with resonances
running inside the loops
and then perform the renormalization.
 If the scalar tadpole is conveniently cancelled out in the way explained
in Sec.~\ref{sec.scalar-tadpole},  the renormalized matrix element
that provides $F_\phi$ is then determined
by the two 1-Particle-Irreducible (1PI) vertex functions
depicted in Fig.~\ref{fig.1PI-vertices}.
More explicitly, we plot in  Figs.~\ref{fig.se} and~\ref{fig.fphi}
the precise diagrams which will be relevant in our R$\chi$T computation
of the pNGB decay constant  up to NLO of $1/N_C$.
Hence, the expression for the physical decay constant consist of two pieces
\begin{equation}\label{deffphi2}
F_{\phi}\,\,\,  = \,\,\,  Z_{\phi}^{\frac{1}{2}}\,\, \,  F_\phi^{\rm 1PI}\, ,
\end{equation}
where $Z_\phi$ stands for the wave-function renormalization constant
of the pNGB given by $\phi^{(B)}=Z_\phi^\frac12 \phi^r$
($\phi=\pi, \, K$)  in the on-shell scheme  (Fig.~\ref{fig.1PI-vertices}.b)
and $F_\phi^{\rm 1PI}$     
denotes the contributions from 1PI topologies for the transition between
an axial-vector current and a bare pNGB $\phi^{(B)}$
(Fig~\ref{fig.1PI-vertices}.a).
The wave-function renormalization constant $Z_\phi=1+\delta Z_\phi$
is related to the pNGB self-energy $\Sigma_\phi(p^2)$ through
\bear
Z_\phi &=& \,\,\,
\left(\, 1\, -\,  \Sigma'_\phi \right)^{-1}
\,\,\,  = \,\,\,
\left(\, 1\, -\, \Frac{d \Sigma_\phi (p^2)}{d p^2}\bigg|_{p^2=m_\phi^2}\right)^{-1}
\,.
\eear

For convenience, we will explicitly separate the tree-level and
one-loop  contributions in R$\chi$T,
\bear
Z_\phi &=& \left(\, \Frac{\widetilde{F}^2}{F_0^2}
\,\,\, +\,\,\, \Frac{8 \wwL_4\,  (2 m_K^2+m_\pi^2)}{F_0^2}
\,\,\,+\,\, \, \Frac{8 \wwL_5\, m_\phi^2}{F_0^2}
   \,\,\,+\,\, \, \Frac{8 \widetilde{L}_{11}\, m_\phi^2}{F_0^2}
\,\,\, -\,\,\, \Sigma'_{\phi, \, 1\ell} \, \right)^{-1}\, ,
\label{eq.Zphi-structure}
\\
\Frac{F_\phi^{\rm 1PI}}{F_0} &=&
\Frac{\widetilde{F}^2}{F_0^2}
\,  +\, \Frac{8 \wwL_4\,  (2 m_K^2+m_\pi^2)}{F_0^2}
\,  +\, \Frac{8 \wwL_5\, m_\phi^2}{F_0^2}
    \,\,\,+\,\, \, \Frac{4 \widetilde{L}_{11}\, m_\phi^2}{F_0^2}
\,\,\, +\,\,\, \Frac{F^{\rm  1PI}_{\phi, \, 1\ell}}{F_0}\, ,
\label{eq.1PI-Fphi-structure}
\eear
with the corresponding one-loop corrections $\Sigma'_{\phi, \, 1\ell}$
and $F^{\rm  1PI}_{\phi, \, 1\ell}$.
This yields the physical decay constant
given by
\bear
F_\phi &=&
\,\,\, F_0\,\,\, \bigg(\,\, \,
  \Frac{\widetilde{F}}{F_0}
\,  +\, \Frac{4 \wwL_4\,  (2 m_K^2+m_\pi^2)}{F_0^2}
\,  +\, \Frac{4 \wwL_5\, m_\phi^2}{F_0^2}
\,\,\, +\,\,\, \Frac{F^{\rm  1PI}_{\phi, \, 1\ell}}{F_0}
\,\,\, +\,\,\, \Frac{1}{2} \Sigma'_{\phi, \, 1\ell}
\,\,\, \bigg) \, ,
\label{eq.Fphi-structure}
\eear
where $Z_\phi^\frac12$ has been expanded in this expression, keeping just
the linear contribution in $\delta Z_\phi$ and dropping other terms
$\cO((\delta Z_\phi)^2)$ or higher.
   In particular, we have used $\widetilde{F}/F_0= 1+\cO(N_C^{-1})$
and  dropped terms $\cO\left( (\widetilde{F}/F_0-1)^2\right)$.
  Notice that there is not a uniquely defined way of truncating the NNLO corrections:
for instance, a slightly different numerical prediction  is obtained
if  instead of the {   expression  }  for $F_\phi$ in Eq.~\eqref{eq.Fphi-structure}
one  employs  the NLO result for  $F_\phi^2$, dropping terms $\cO((\delta Z_\phi)^2)$
or higher as we did in Eq.~\eqref{eq.Fphi-structure}.
  The spurious coupling $\widetilde{L}_{11}$
(corresponding to an operator proportional to the equations of motion~\cite{SanzCillero:2009ap})
becomes then cancelled out and disappears  from the physical
observable.

Since we did not consider $\cO(p^6)$ operators
in the Lagrangian $\mL_G^{\rm LO}$ in Eq.~\eqref{lolagrangian},
we will neglect the terms
$\cO(\wwL_{4,5}^2 m_\phi^4/F^4)$ in the decay constant in Eq.~\eqref{eq.Fphi-structure}
which would arise from the expansion at that order of
$Z_\phi^\frac{1}{2}$.
In spite of having the same tree-level structure,
in general the loop UV--divergences in $Z_\phi$ and
$F_\phi^{\rm 1PI}$, given respectively in Eqs.~\eqref{eq.Zphi-structure}
and~\eqref{eq.1PI-Fphi-structure}, are different.
Thus, one must combine
these two quantities into Eq.~\eqref{eq.Fphi-structure} in order to get
a finite decay constant $F_\phi$ by means of the renormalization
of $\widetilde{F}$, $\wwL_4$ and~$\wwL_5$.

In the $N_C\to\infty$ limit, the meson loops are absent one has
\bear
\wwL_4 \,  
=\, \widetilde{L}_4    
\, , \qquad\qquad  \wwL_5 \,     
\, = \, \left(\widetilde{L}_5 +\Frac{c_d c_m}{M_S^2}\right)   
\, =\, \Frac{F_0^2}{4 M_S^2}\, ,
\eear
  where we considered  the large--$N_C$ high-energy constraints
 $\widetilde{L}_4  
= \widetilde{L}_5   
=0$ and  $(4 c_d c_m/F_0^2)   
=1$ from the scalar form-factor~\cite{Jamin:2001zq}.
At LO in $1/N_C$ this yields the  prediction~\cite{SanzCillero:2004sk}
\bear
F_\phi&=& \,\,\,
F_0\,\,\,\bigg(\,\,\, 1\,\,\, +\,\,\, \Frac{4 c_d c_m }{F_0^2} \, \Frac{m_{\phi}^2}{M_S^2}
\,\,\, \bigg)
\,\,\, = \,\,\,
F_0\,\,\,\bigg(\,\,\, 1\,\,\, +\,\,\, \Frac{m_\phi^2}{M_S^2}
\,\,\, \bigg)\, ,
\label{eq.large-Nc-Fphi}
\eear
  which   reproduce the  $F_\pi$ and $F_K$ lattice data fairly well
up to pion masses of the order of 700~MeV~\cite{SanzCillero:2004sk}.
The large--$N_C$ relation
$c_d=c_m$~\cite{Jamin:2001zq} was used in Ref.~\cite{SanzCillero:2004sk}
to produce Eq.~\eqref{eq.large-Nc-Fphi}, where it led to the  relation
$m_\phi^2= B_0 (m_{q_1} +m_{q_2})$ between the pNGB mass and the masses of its two
valence quarks.
We have also used that $\widetilde{F}/F_0=1$ and
$\Sigma_{\phi,1\ell}=F^{\rm 1PI}_{\phi,1\ell}=0$ when  $N_C\to\infty$.
The only region where this description deviated significantly from the
data was in the light pion mass range, where the chiral logs
need to be included to properly reproduce the lattice simulation
in that regime~\cite{Davies:2003fw,Davies:2003ik}.
   Here in Eq.~\eqref{eq.large-Nc-Fphi} the coupling
$F_0$  implicitly refers to the $n_f=3$ decay constant in that same limit,
this is,  at large $N_C$.

  In summary,  our calculation of the pNGB decay constants
$F_\phi$ (with $\phi=\pi,\, K$) is sorted out in the form
\bear
\Frac{F_\phi}{F_0} \quad &=&\quad  \Frac{\Delta F_\phi}{F_0}\bigg|_{\cO(N_C^0)}\quad +\quad
\Frac{\Delta F_\phi}{F_0}\bigg|_{\cO(N_C^{-1})}\quad +\quad   ...
\eear
with the dots standing for terms
of $\cO(N_C^{-2})$ and higher, which will be neglected in the present article.
In the joined  large--$N_C$ and chiral limits one has the right-hand side becomes equal to one
by construction, as  $F_0=\lim_{m_{u,d,s}\to 0} F_\phi$.
At large--$N_C$, the relevant couplings in the quark mass corrections to $F_\phi$
are related with the scalar form-factor and can be fixed through
high-energy constraints~\cite{SanzCillero:2004sk,Jamin:2001zq}.
However, one should be aware that it is not possible to have a full control of the
quark mass corrections beyond the linear $m_q$ term.
   In fact,  including all possible $m_q^2$ corrections  corresponds
to considering  the full sets of local $\cO(p^6)$ operators
without resonance fields.
The complexity of higher order $m_q$ corrections not only happens
for the NLO in $N_C$ but also for the LO case.
For instance, large--$N_C$ contributions  to the scalar (vector) multiplet mass splitting
can be in principle of an arbitrary order in $m_q$,
leading to LO (NLO) corrections in $1/N_C$ to $F_\phi$ with arbitrary powers of the quark mass.
Clearly, there is not a uniquely defined  truncation procedure.

 \subsection{Renormalization in R$\chi$T}

The calculation of the Feynman diagrams contributing to $F_\phi$
up to NLO in $1/N_C$  is straightforward  (Figs.~\ref{fig.se} and \ref{fig.fphi}),
though the final results are quite lengthy  and have been relegated to
App.~\ref{app.diagrams} for the sake of clarity and in order not to interrupt our discussion.
{  We take into account the $m_q$ dependence of the resonance masses
in the propagators in the loops, which is given by  Eq.~\eqref{masssplit}.  }

{   In order to have finite results for the physical quantities $F_\pi$ and $F_K$,
the next step consists on performing  the renormalization.  }
As in conventional $\chi$PT~\cite{gl845}
we use the dimensional regularization method and
the $\rm{\overline{MS}}-1$ renormalization scheme
where we will subtract from the Feynman integrals the UV--divergence
\begin{eqnarray}
\frac{1}{\hat\epsilon}&=&
 \mu^{-2\epsilon}  \,
\left( \frac{1}{\epsilon}-\gamma_E + \ln{4\pi}+1\right)
\,\,\, = \,\,\,
 \frac{1}{\epsilon}-\gamma_E + \ln{4\pi}+1 -\ln\mu^2+\cO(\epsilon)
\,,
\quad \quad (\epsilon= 2-\frac{D}{2})\,.
\nn\\
\end{eqnarray}
The UV--divergences from loops can be absorbed
through a convenient renormalization of the R$\chi$T couplings
$C_\chi=\widetilde{F},\, \wwL_4,\, \wwL_5$ in the form
\be
C_\chi\,\,\, =\,\,\, C_\chi^r(\mu)\,\,\, +\,\,\, \delta C_\chi(\mu)\, ,
\ee
where the $C_\chi^r(\mu)$ are the finite renormalized couplings
and the counter-terms  $\delta C_\chi(\mu)$  are infinite and
cancel out the one-loop UV--divergences.
The $\rm{\overline{MS}}-1$ scheme is usually employed in $\chi$PT and R$\chi$T,
where the subtracted divergence is of the form
\bear
\delta C_\chi(\mu)&=& \, -\, \Frac{\Gamma^{C_\chi}}{32\pi^2}\, \,
\Frac{1}{\hat{\epsilon}}\, ,
\eear
and the renormalized coupling has a renormalization group running given by
\bear
\Frac{d C_\chi (\mu)}{d\ln\mu^2} &=& \, -\, \Frac{\Gamma^{C_\chi}}{32\pi^2}\, .
\eear

This will be the scheme considered to renormalize $\wwL_4$
and $\wwL_5$ in this article.
More precisely, the renormalization of
$\widetilde{\widetilde{L}}_4=\widetilde{L}_4$
and $\widetilde{\widetilde{L}}_5=\widetilde{L}_5 \, +\,c_d c_m/M_S^2$
is given by
\begin{eqnarray}
\Gamma^{\widetilde{\widetilde{L}}_4} \,\,\,&=& \,\,\, \frac{1}{8}\,
\bigg[ \, 1\, +\, \frac{4 c_d c_m}{F_0^2}\, +\,
\frac{2c_d^2}{F_0^2}\, (1-4e_m^S)\,
-\frac{3G_V^2}{F_0^2}\, (1-4 e_m^V) \bigg]\,,\nonumber \\
\Gamma^{\widetilde{\widetilde{L}}_5} \,\,\,  &=&\,\,\,
\frac{3}{8}\,
\bigg[ \, 1\, -\, \frac{4 c_d c_m}{F_0^2}\, +\,
\frac{2c_d^2}{F_0^2}\, (1-4e_m^S)\,
-\frac{3G_V^2}{F_0^2}\, (1-4 e_m^V) \bigg]\,.
\end{eqnarray}
One may compare this result with that in $SU(3)$ $\chi$PT,
$\Gamma^{L^{\chi PT}_5} = 3\, \Gamma^{L^{\chi PT}_4} =3/8$~\cite{gl845}.

{  The $\overline{MS}-1$
one-loop renormalization  $\delta \widetilde{F}$ is found to be   }
\begin{eqnarray}
\frac{\delta \widetilde{F}        
}{F_0} &=&-\frac{1}{16\pi^2}\bigg( \frac{3c_d^2 M_S^2}{F_0^4}
+\frac{2c_d^2 M_0^2}{3F_0^4}-\frac{9G_V^2 M_V^2}{2F_0^4} \bigg) \,\,\,
 \Frac{1}{\hat{\epsilon}}  \,.
\end{eqnarray}
This result recovers the scalar and vector resonance contributions obtained
in Ref.~\cite{SanzCillero:2009ap}.
Though the $M_0$ term in the previous equation is in principle
$1/N_C$ suppressed, it can be important in the phenomenological discussion
as its numerical value of $M_0$ is not small.
Due to the inclusion of the heavier resonance states and the singlet $\eta_1$,
the renormalization in R$\chi$T is a bit different from the conventional one in $\chi$PT with only pNGB.
Indeed, it resembles a bit the situation in Baryon $\chi$PT,
where the loops generate power-counting breaking terms which contribute
at all orders in the chiral expansion~\cite{PCB-terms}.
  For instance, based on dimensional analysis~\cite{Weinberg:1979}
one can prove that the $\cO(p^2)$ coupling $F_0$ does not get renormalized
at any order in $\chi$PT since any possible loop correction is always $\cO(p^4)$ or higher.
This is not the case in R$\chi$T, where in general  one needs to renormalize the couplings of the
LO Lagrangian to cancel out the one-loop UV-divergences~\cite{L9a,SanzCillero:2009ap,RChT-RGE}.
Moreover, though subleading in $1/N_C$,  the R$\chi$T loops with massive states generate
power-counting breaking terms from the point of view of the $\chi$PT chiral counting,
in the same way as it happens in Baryon $\chi$PT~\cite{PCB-terms}.
We will explicitly see in the next section that,
the matching of the R$\chi$T and $\chi$PT results  in the low-energy region
fixes completely the LO coupling $\widetilde{F}$ and solve the problem with
the power-counting breaking terms.

Notice that in the present work, after the renormalizations of
$\widetilde{F}, \wwL_4$ and $\wwL_5$, we obtain a finite
result for our physical observables  $F_\pi$ and $F_K$.
{  In other words,    all the one-loop UV divergences of
the pion and kaon decay constant calculation    }
can be cancelled out through the     convenient
renormalizations $\delta \widetilde{F}, \delta\wwL_4$ and $\delta \wwL_5$.

\subsection{Matching  R$\chi$T and $\chi$PT}

In order to establish the relation between the $\chi$PT LECs and
{   the couplings   }    from R$\chi$T,
it is necessary to perform the chiral expansion of the decay constants calculated in R$\chi$T and then match with the pure $\chi$PT results.
This procedure resembles the reabsorption of the power breaking terms into the lower order couplings
in Baryon $\chi$PT~\cite{PCB-terms}.
In such a way, we can relate the $\chi$PT LECs with the resonance couplings
including not only the { leading order contributions in $1/N_C$  }
but also the $1/N_C$ corrections.

The pNGB decay constants are given in $SU(3)$ $\chi$PT up to $\cO(p^4)$  by Ref.~\cite{gl845}
\bear
F_\pi&=&
\,\,\, F_0\,\,\, \left(\,\, \,
1
\,  +\, \Frac{4 L^{\chi PT}_4\,  (2 m_K^2+m_\pi^2)}{F_0^2}
\,  +\, \Frac{4 L^{\chi PT}_5\, m_\pi^2}{F_0^2}
\,\,\, -\, \Frac{1}{2 F_0^2}\,
\left[ 2 i A_0(m_\pi^2)\,  +\,  i A_0(m_K^2) \right]
\,\,\, \right) \, ,
\nn\\
F_K &=&
\,\,\, F_0\,\,\, \left(\,\, \,
1
\,  +\, \Frac{4 L^{\chi PT}_4\,  (2 m_K^2+m_\pi^2)}{F_0^2}
\,  +\, \Frac{4 L^{\chi PT}_5\, m_K^2}{F_0^2}
\,\,\, -\, \Frac{3}{8F_0^2}\,
\left[ i A_0(m_\pi^2)\,  +\, 2 i A_0(m_K^2) \,  + \,i A_0(m_{\eta_8}^2)\right]
\,\,\, \right) \, ,
\nn\\
\label{eq.ChPT-Fphi}
\eear
with $m_{\eta_8}^2 = (4 m_K^2 -m_\pi^2)/3$ and
the one-point Feynman integral $A_0(m^2)$ is given in Appendix~\ref{app.Feynman-int}
whose  UV--divergences
are conveniently renormalized through
$L_4^{\chi PT}$ and $L_5^{\chi PT}$.

  The chiral singlet pNGB $\eta_1$ requires a particular treatment.
When the chiral expansion of the R$\chi$T { expressions  }
is performed to match $SU(3)$--$\chi$PT,
we do not take the singlet $\eta_1$ mass $M_0$ as a small expansion parameter.
Instead, we keep its full contribution in spite of being its effect  suppressed by $1/N_C$.
It is known that in the low-energy EFT where the $\eta'$ has been integrated
these contributions may become phenomenologically important~\cite{Kaiser:2000gs}.
Expanding the decay constants in R$\chi$T in powers of $m_\phi^2$   
and then matching with  $SU(3)$--$\chi$PT up to $\cO(p^4)$,
we obtain the following relations
\begin{eqnarray}
 1
&=&
 \,\,\, \frac{\widetilde{F}^r(\mu)}{F_0}
\nn\\
&&  \,\,\,  + \,\,\,
\frac{1}{16\pi^2}\bigg[ \frac{c_d^2 M_S^2 }{F_0^4}
\big( \frac{7}{6} - \frac{7}{3}\ln{\frac{M_S^2}{\mu^2}} \big)
+\frac{c_d^2}{3F_0^4}\frac{M_S^4-M_0^4+2M_0^4\ln{\frac{M_0^2}{\mu^2}}-2M_S^4\ln{\frac{M_S^2}{\mu^2}}}{M_S^2-M_0^2}
\nonumber\\
&&\qquad\qquad
+\frac{G_V^2 M_V^2}{F_0^4}  \big( \frac{3}{4} + \frac{9}{2}\ln{\frac{M_V^2}{\mu^2}}
\big) \bigg]\,,
\label{eq.Ftilde-matching}
\\
\nn\\
  L_4^{\rm{\chi PT}, r}(\mu)
&=&
  \widetilde{\widetilde{L}}_4^r(\mu)
\nn\\
&& \,\,\, +\, \,\,
\frac{1}{16\pi^2 F_0^2}\,\,\bigg\{
\frac{c_d^2}{144(M_S^2-M_0^2)^2}\bigg[ (M_0^2-9M_S^2)(M_0^2-M_S^2)+8M_0^4\ln{\frac{M_0^2}{\mu^2}}
\nonumber \\& &\qquad \qquad \qquad \qquad \qquad \qquad \qquad \qquad
-2(13M_0^4-18M_0^2M_S^2+9M_S^4)\ln{\frac{M_S^2}{\mu^2}} \bigg]
\nn\\
&& +\frac{1}{8}c_d c_m-\frac{1}{4}c_d c_m\ln{\frac{M_S^2}{\mu^2}} +\frac{1}{32}G_V^2+\frac{3}{16}G_V^2\ln{\frac{M_V^2}{\mu^2}}
\nonumber \\& &
+\,c_d^2 e_m^S(\frac{1}{4}+\frac{1}{2}\ln{\frac{M_S^2}{\mu^2}})-G_V^2 e_m^V(\frac{7}{8}+\frac{3}{4}\ln{\frac{M_V^2}{\mu^2}}) \bigg\}\,,
\label{l4chptmatching}
\\
\nn\\
L_5^{\rm{\chi PT},r}(\mu)
&=&
   \widetilde{\widetilde{L}}_5^r(\mu)
\nn\\
&&
+ \frac{1}{16\pi^2 F_0^2}\,\bigg\{ \frac{c_d^2}{48(M_0^2-M_S^2)}\bigg[
9(M_0^2-M_S^2)-16M_0^2\ln{\frac{M_0^2}{\mu^2}}
-2(M_0^2-9M_S^2)\ln{\frac{M_S^2}{\mu^2}} \bigg] +
\nonumber \\& &
\frac{c_d^2 e_m^S}{12(M_S^2-M_0^2)^2}\bigg[ (M_0^2-9M_S^2)(M_0^2-M_S^2)
\nn\\
&&\qquad\qquad \qquad\qquad
+8M_0^4\ln{\frac{M_0^2}{\mu^2}}
+2(5M_0^4-18M_0^2M_S^2+9M_S^4)\ln{\frac{M_S^2}{\mu^2}} \bigg]
\nonumber \\& &
-\frac{3}{8}c_d c_m+\frac{3}{4}c_d c_m\ln{\frac{M_S^2}{\mu^2}} +\frac{3}{32}G_V^2+\frac{9}{16}G_V^2\ln{\frac{M_V^2}{\mu^2}}
-G_V^2 e_m^V(\frac{21}{8}+\frac{9}{4}\ln{\frac{M_V^2}{\mu^2}}) \bigg\}\,.
\label{l5chptmatching}
\end{eqnarray}
We have matched the chiral expansion of our R$\chi$T predictions for $F_\phi$
in powers of the quark masses $m_q$ [on the right-hand side
of Eqs.~\eqref{eq.Ftilde-matching}--\eqref{l5chptmatching}]
to  the corresponding chiral expansion in $\chi$PT [left-hand side
of Eqs.~\eqref{eq.Ftilde-matching}--\eqref{l5chptmatching}].
Eq.~\eqref{eq.Ftilde-matching} stems from the matching at $\cO(m_q^0)$
and Eqs.~\eqref{l4chptmatching} and~\eqref{l5chptmatching} are derived from the
chiral expansion at $\cO(m_q^1)$.
{   The one-loop contributions in Eq.~\eqref{eq.ChPT-Fphi}
are exactly matched and one recovers the  correct running
for  the $L_4^{\chi PT}(\mu)$  and $L_5^{\chi PT}(\mu)$ predictions.   }
Notice that if we took the on-shell renormalization
scheme from Ref.~\cite{SanzCillero:2009ap}
instead of the $\rm{\overline{MS}}-1$ scheme considered here
Eq.~\eqref{eq.Ftilde-matching} would become
$1=\widetilde{F}^r/F_0$.

The final renormalized expression for the pNGB decay constants is
\begin{eqnarray}\label{fphifinal}
 F_{\phi} \,\,\,=\,\,\,  F_0\, \left[ \,  1
 + \frac{4\widetilde{\widetilde{L}}_4^r (2m_K^2+m_\pi^2)}{F_0^2}
 + \frac{4 \widetilde{\widetilde{L}}_5^r m_{\phi}^2}{F_0^2}
\,\,\, + \,\,\, \Frac{F_{\phi, 1\ell}^{\rm 1PI,r}}{F_0}  + \frac{1}{2} \Sigma'^{r}_{\phi, 1\ell}
  + \bigg(   \frac{\widetilde{F^r}}{F_0} -1\bigg)     
 \, \right]\,,
\end{eqnarray}
     with last term $\left(\frac{{\widetilde{F^r}}}{F_0} -1\right)$
given by the matching condition in Eq.~\eqref{eq.Ftilde-matching}.
This  ensures that the  renormalized contributions from the one-loop diagrams
are appropriately cancelled out in the chiral limit so that the decay constants $F_\phi$ become
equal to $F_0$  when $m_q\to 0$.
In the same way,  in our later analysis,
 $\widetilde{\widetilde{L_4}}^r$ and $\widetilde{\widetilde{L_5}}^r$
 will be expressed in terms of
$L_4^{\chi PT, r}$ and $L_5^{\chi PT, r}$, respectively,
by means of  the $\chi$PT matching relations in Eqs.~\eqref{l4chptmatching}
and \eqref{l5chptmatching}.
This will allow us to deal in a more direct and convenient way
with  the $\chi$PT LECs in our R$\chi$T predictions.
   Our theoretical predictions for $F_\phi$ will depend only on
\bear
\mbox{\bf Tree-level contributions:}&&
\qquad \widetilde{F}^r\,\,\, \{\to F_0\}\, ,\qquad
\wwL_4^r \,\,\, \{\to L_4^{\chi PT,r} \}\, , \quad
\wwL_5^r\,\,\, \{ \to L_5^{\chi PT, r} \}\, ,
\qquad \label{eq.tree-couplings}
\\
\mbox{\bf One-loop contributions:}&&
\qquad e_m^V\, ,\quad e_m^S\, , \quad c_m\, ,
\qquad  \label{eq.loop-couplings}
\\
&&\qquad c_d\, ,\quad G_V\, ,\qquad\qquad
\nn\\
&&\qquad M_V\, ,\quad M_S\, , \quad M_0\, .  \qquad
\nn
\eear
The  tree-level contributions from $\widetilde{F}$, $\wwL_4^r(\mu)$ and $\wwL_5^r(\mu)$
will be expressed in terms of $F_0$, $L_4^{\chi PT,r}(\mu)\, , L_5^{\chi PT, r}(\mu)$
in the phenomenological analysis.
As we are carrying our decay constant computation up to   NLO in $1/N_C$,
the LECs of $F_0$, $L_4^{\chi PT,r}(\mu)\, , L_5^{\chi PT, r}(\mu)$ correspond to the renormalized couplings at that order,
not just their large--$N_C$ values.
In contrast, the remaining parameters only appear within loops. They do not get renormalized at this order
and correspond to their large--$N_C$  values.
In our fits to lattice simulations we will always fit the parameters in the first line of
Eq.~\eqref{eq.loop-couplings}, we will use high-energy constraints for those in the second line
(although we will also check the impact of fitting $G_V$ or setting it to particular values),
and the parameters in the third line of Eq.~\eqref{eq.loop-couplings}
will be always introduced as inputs.

   We will also analyze the lattice results for the ratio $F_K/F_\pi$. Following
the principle considered before, we will fit the data with our theoretical prediction
expanded up to NLO:
\bear
\Frac{F_K}{F_\pi} &=&
 \,  1 \,\,\,+\,\,\, \frac{4 \widetilde{\widetilde{L}}_5^r \, (m_K^2-m_\pi^2)}{F_0^2}
\,\,\, + \,\,\, \left(\Frac{F_{K, 1\ell}^{\rm 1PI,r}}{F_0}
+ \frac{1}{2} \Sigma'^{r}_{K, 1\ell}\right)
\,\,\, -  \,\,\, \left(\Frac{F_{\pi, 1\ell}^{\rm 1PI,r}}{F_0}
+ \frac{1}{2} \Sigma'^{r}_{\pi, 1\ell}\right)
\, . \label{eq.expanded-ratio}
\eear

Apart from  the present analysis of lattice data,
Eqs.~\eqref{l4chptmatching} and~\eqref{l5chptmatching}
can be also employed to predict the $L_4^{\chi PT}$ and $L_5^{\chi PT}$
chiral LECs  in terms of resonance parameters.
These NLO expressions fully recover  the one-loop running of the LECs
and can be used to extract the chiral couplings at any renormalization
scale $\mu$. Furthermore, by imposing high-energy constraints
in the way previously considered in analogous one-loop
analyses~\cite{SanzCillero:2009ap,Rosell-L8,Rosell-L9-L10}, it should be possible to provide
similar NLO predictions in $1/N_C$  in terms of $F_0$ and the resonance
masses $M_R$.


\section{Phenomenological discussions}
\label{sec.pheno}

\subsection{Inputs and constraints}

As previously mentioned in Introduction, we confront our theoretical
calculation of the pNGB decay constants to the {  lattice data from  }  different lattice
collaborations~\cite{Davies:2003fw,Davies:2003ik,Aoki:2010dy,Arthur:2012opa,Durr:2010hr}.

\begin{figure}
\begin{center}
 \includegraphics[angle=0,width=1.0\textwidth]{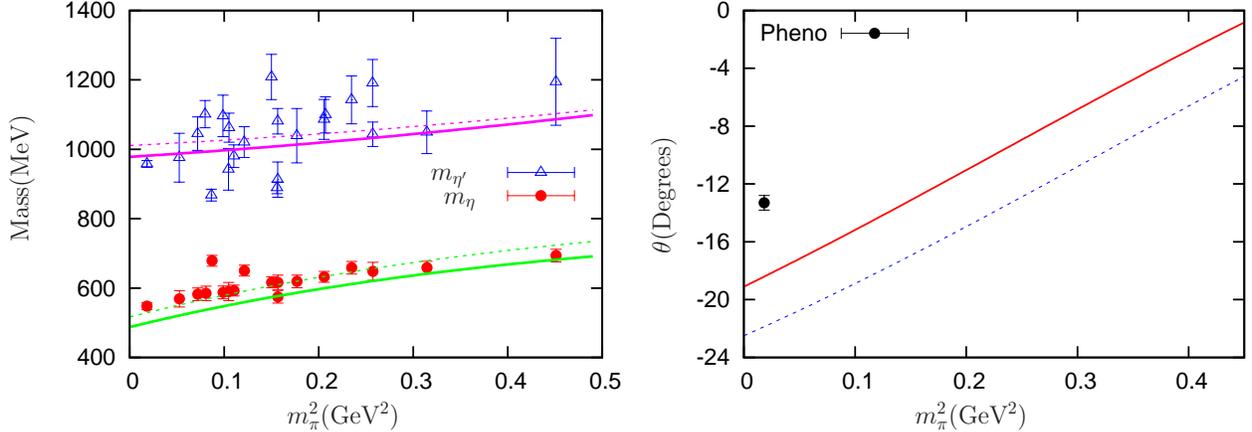}
 \caption{{\small  Masses and mixing angle of $\eta$ and $\eta'$.
 The points in the left panel are taken from Ref.~\cite{Michael:2013vba},
 which summarizes the data from ETM~\cite{Michael:2013gka}, RBC-UKQCD~\cite{Christ:2010dd},
 HSC~\cite{Dudek:2011tt} and UKQCD~\cite{Gregory:2011sg} collaborations.
The mixing angle $\theta=(-13.3\pm 0.5)^\circ$ extracted in the phenomenological
analysis~\cite{theta-exp} for physical masses     
is  plotted in the right panel.
The solid lines are obtained   by using
the physical strange quark mass ($m_s=m_{s,{\rm Phys}}$)
and the dashed lines come from  employing  $m_s= 1.2 \,m_{s,{\rm Phys}}$.
The value    $M_0=850$~MeV is taken as input~\cite{Feldmann:1999uf}.
We remind the reader this plot is a prediction,  not a fit.
 }}\label{fig.meta}
\end{center}
\end{figure}

  In addition, we also take into account the lattice determination
of the $\rho(770)$ mass with
varying quark masses~\cite{Lang:2011mn,Aoki:2011yj,Pelissier:2012pi,Dudek:2012xn},
which helps us {  to  } constrain the vector mass splitting coupling $e_m^V$ in Eq.~\eqref{lagemr}.
For the scalar resonances, as we mentioned previously, we only consider those that survive
at  large $N_C$. In fact, this is not a settled problem yet.
For example, the Inverse Amplitude Method analyses~\cite{RuizdeElvira:2010cs}
found that the $f_0(500)$ or $\sigma$ resonance could fall {  down  }  to the real axis at large values of $N_C$, meaning that it survives
as a conventional $\bar{q}q$ state   {   at  large $N_C$,  }  while the $f_0(980)$ disappears in that limit.
In the N/D approach, the situation is just the opposite~\cite{Guo:2011pa,Guo:2012ym,Guo:2012yt}.
However it is interesting to point out that though the resonance trajectories for $N_C=3$ to $\infty$ are quite different in the two approaches,
there is an important common {  conclusion: a scalar }   resonance with mass around 1~GeV at large $N_C$ is necessary to fulfill the
semi-local duality in $\pi\pi$ scattering. In Refs.~\cite{Guo:2011pa,Guo:2012ym,Guo:2012yt}, it was also proved that the 1~GeV scalar resonance
at large $N_C$ is needed to satisfy the Weinberg sum rules in the scalar and pseudoscalar sectors.
Therefore it  { seems proper}  to set the bare scalar resonance mass
at large $N_C$ around 1~GeV.
{  This  } is also supported by our previous analysis in Ref.~\cite{Guo:2009hi}.
In the following we take the result
   $M_S=980$~GeV from Ref.~\cite{Guo:2009hi} {as an input  while
the value of the scalar mass splitting coupling $e_m^S$ will  be fitted    }
in this work, as the value in the previous reference is determined
with too large error bars.

The leading order  expressions have been  employed
in  our theoretical analysis to relate the squared kaon mass  with the varying  squared pion mass
and $m_{u/d}$:
\begin{eqnarray}
m_\pi^2 &=& 2 B \, m_{u/d}\, ,
\label{eq.mpi}
\\
m_K^2&=& B\, (m_s+m_{u/d})\, =\,
\bigg( m_{K,{\rm phys}}^{2}-\frac{m_{\pi,{\rm phys}}^{2}}{2} \bigg)\alpha_{m_s}  + \frac{m_\pi^2}{2}\,,
\label{mkmpi}
\end{eqnarray}
where $m_{K,{\rm phys}}$ and $m_{\pi,{\rm phys}}$ denote the physical masses of kaon and pion.
   Different values of $\alpha_{m_s}=m_s/m_{s,{\rm phys}}$
correspond  in this equation  to the situations with different strange quark masses
whereas $\alpha_{m_s}=1$ refers to  the physical $m_s$ case.
We will always take $\alpha_{m_s}=1$ in all the fits  in this article,
considering only  lattice simulation data with $m_s=m_{s,{\rm phys}}$.
Later on, after performing the fit, we will study to what extent
our results depend on the linear quark mass relations
in Eqs.~\eqref{eq.mpi}~and~\eqref{mkmpi}
by varying $\alpha_{m_s}$.
Indeed, this insensitivity to higher order corrections was already observed for
$m_\pi^2/(2 B m_{u/d})$~\cite{Mpi-lattice,Durr:2010hr}.
This ratio was found to show a very small dependence on $m_{u/d}$ in the whole range of values
of the simulation~\cite{Mpi-lattice,Durr:2010hr}, supporting the
description given by Eqs.~\eqref{eq.mpi} and~\eqref{mkmpi} and used in this article.

    Since the $\eta$ and $\eta'$ only enter the expressions of $F_\pi$ and $F_K$
through the chiral loops, it is enough to consider the
  leading order mixing produced by the Lagrangian~\eqref{lolagrangian}
for their masses and mixing angle (see App.~\ref{app.eta-etap} for details).
The chiral limit of the singlet $\eta_1$ mass ($M_0$) is by definition independent
of the light quark mass  and will  take the fixed value $M_0=850$~MeV
in this article~\cite{Feldmann:1999uf}.
In Fig.~\ref{fig.meta} one can see the fair  agreement of the LO prediction with lattice
simulations~\cite{Michael:2013vba,Michael:2013gka,Christ:2010dd,Dudek:2011tt,Gregory:2011sg}
and previous phenomenological analyses~\cite{theta-exp}   for the physical quark mass.     The one-parameter  fit to  lattice
data~\cite{Michael:2013vba,Michael:2013gka,Christ:2010dd,Dudek:2011tt,Gregory:2011sg}
for $m_\eta$ and $m_{\eta'}$ (Fig.~\ref{fig.meta})
yields essentially the same value ($M_0\simeq 835$~MeV),
very close to the input $M_0=850$~MeV which  will be employed  all through the paper
and indistinguishable {   in  } Fig.~\ref{fig.meta} when plotted.

    In Fig.~\ref{fig.meta}, the solid lines correspond to our predictions with $\alpha_{m_s}=1$ and
the dashed lines refer to  the case with $\alpha_{m_s}=1.2$. It is clear that the change caused by using different strange quark
masses in $\eta-\eta'$ mixing is mild.
On the other hand, it is remarkable that the leading order mixing from
{   $U(3)-\chi$PT    }
can reasonably reproduce the lattice simulation data
for the masses of $\eta$ and $\eta'$, as shown in the left panel of Fig.~\ref{fig.meta}.
In right panel, we show the leading order mixing angle $\theta$ with varying pion masses, i.e.
with varying light $u/d$ quark masses. As expected, when the $u/d$ quark mass approaches to the strange quark mass,
i.e. the pion mass tends to the kaon mass, there is no mixing between $\eta_1$ and $\eta_8$, as their
mixing strength is proportional to the $SU(3)$ breaking $m_K^2-m_\pi^2$.
Likewise,  this result  gives extra support to the linear dependence on the light quark masses
for  $m_\phi^2$ assumed in Eqs.~\eqref{eq.mpi} and~\eqref{mkmpi}
as an approximation in this article.

   In the fit,  we will use the chiral limit mass of the vector resonance multiplet
computed in Ref.~\cite{Guo:2009hi} as an input:
\begin{eqnarray}
\quad M_V=764.3~{\rm MeV} \,.
\end{eqnarray}

 Imposing   the high energy constraints dictated by QCD is an efficient way
to reduce the free couplings in effective field theory. In addition it makes the
effective field theory inherit more properties from QCD. In R$\chi$T literature,
it is indeed quite popular to constrain the resonance couplings
through the high energy behaviors {   of   }
form factors~\cite{Rosell-L9-L10,Jamin:2001zq},
meson-meson scattering~\cite{Guo:2007ff,Guo:2011pa},
Green functions~\cite{SanzCillero:2009ap},
tau decay   form-factors~\cite{Guo:2008sh,Guo:2010dv},  etc.
Among the various constraints obtained in literature,
two of them are relevant to our current work
\begin{eqnarray}\label{hecd}
c_d &=&   {     \frac{F_0^2}{4c_m}\,,     }
\\
G_V &=&    {     \sqrt{\frac{F_0^2-2c_d^2}{3}}\,,    }
\label{hegv}
\end{eqnarray}
resulting from the analyses of the scalar form factor~\cite{Jamin:2001zq}
and partial-wave $\pi\pi$ scattering~\cite{Guo:2007ff} at large $N_C$, respectively.

The renormalization scale $\mu$ will be set
   at 770~MeV,  corresponding   the renormalized  LECs determined later
to their values at that scale.

\subsection{Fit to lattice data}
\label{sec.bestfit}

We use the CERN MINUIT package to perform the fit.
The values  of the six free parameters from the fit read
\begin{eqnarray}\label{fitresult}
&&  F_0=80.0\pm1.0~{\rm MeV}\,, \qquad
L_4^{\chi PT}=(-0.11\pm0.06)\times 10^{-3}~\,,  \qquad
L_5^{\chi PT}=(0.59\pm0.08)\times 10^{-3}\,,
\nn\\
&&c_m=54.5\pm3.3~{\rm MeV}\,, \qquad
e_m^V=-0.236\pm0.005\,,  \qquad
e_m^S=-0.204\pm0.024\,,
\end{eqnarray}
with $\chi^2/$d.o.f$=90.8/(52-6)$.
  The strange quark mass is kept fixed to $m_{s,{\rm phys}}$ in this fit.
We point out that one should  take the value of $\chi^2$ from the fit
as a mere orientation of the goodness of the fit rather than in its precise
statistical sense: lattice simulation results should not be taken as real experimental data
for various quark masses as they are in general
highly correlated and systematic uncertainties should be also properly accounted.
This gets even worse when combining data from different groups.
For a detailed discussion see Ref.~\cite{Durr:2010hr}.
The aim of this work is to provide a first quantitative analysis
of the potentiality of these type of hadronic observables, i.e. $F_\pi$ and $F_K$,
for the study of resonance properties.

  By substituting the results from  Eq.~\eqref{fitresult} in
the high energy constraints given in Eqs.~\eqref{hecd} and \eqref{hegv} one gets
\begin{eqnarray}\label{cdgv}
 c_d=29.3\pm1.9~{\rm MeV}\,, \quad G_V= 39.5\pm0.9~{\rm MeV}\,.
\end{eqnarray}

  The negative values for $e_m^R$ indicate that the resonance masses
  grow with $m_q$ as one can see from Eq.~\eqref{masssplit}.
They  are found in agreement with the previous   estimates
$e_m^V=-0.228\pm 0.015$ and $e_m^S=-0.1\pm 0.9$~\cite{Guo:2009hi}.
The present determinations for $c_d$ and $c_m$ are compatible with those in Ref.~\cite{Guo:2009hi}: $c_d=26\pm 7$~MeV
and $c_m=80\pm 21$~MeV.
Nonetheless, we find large discrepancy for the value of the $\rho-\pi\pi$ coupling given in Ref.~\cite{Guo:2009hi}: $G_V=63.9\pm 0.6$~MeV.
The reason for the large discrepancies of the $G_V$ values will be analyzed in detail in next section.

\begin{figure}
\begin{center}
 \includegraphics[angle=0,width=1.0\textwidth]{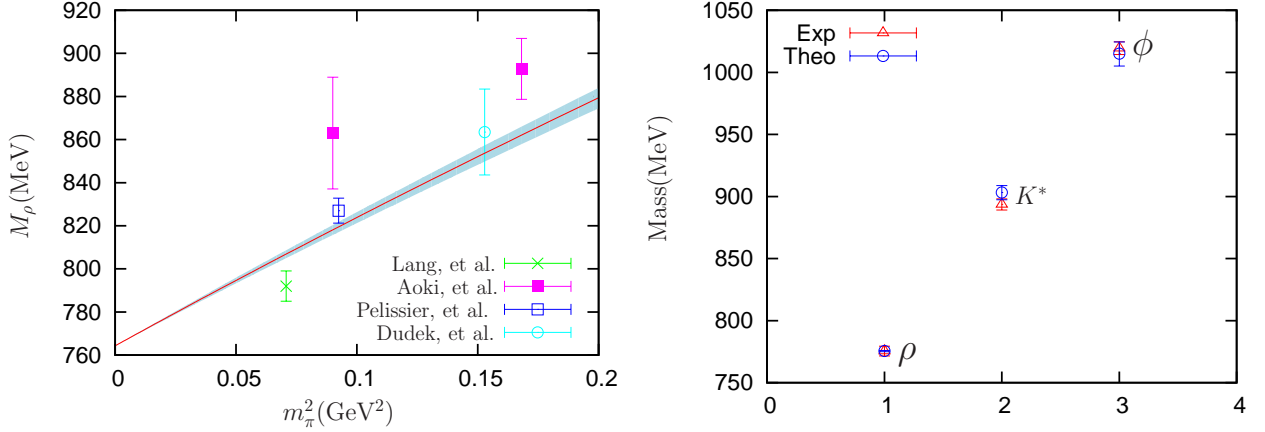}
 \caption{{\small  Fit results for the vector masses. The left panel
 { shows  }  the square pion mass dependence of $M_\rho$.
 {  The   }   lattice data in this panel are taken from Refs.~\cite{Lang:2011mn,Aoki:2011yj,Pelissier:2012pi,Dudek:2012xn}.
 {  The right panel shows the  } masses of $\rho(770)$, $K^*(892)$ and $\phi(1020)$
 with the physical pion mass. The shaded area in the left panel and the empty circles
in the right panel represent our estimation of the error bands,
which are explained in detail in the text.
}}\label{fig.mrho}
\end{center}
\end{figure}

In the left panel of Fig.~\ref{fig.mrho}, we show our fit results together
with the lattice data for $M_\rho$ with different
pion masses, which are originally taken from
Refs.~\cite{Lang:2011mn,Aoki:2011yj,Pelissier:2012pi,Dudek:2012xn}.
Due to the large error bars of these data, the stringent constraint
on the vector mass splitting parameter
$e_m^{V}$ {comes}  from the determination of physical masses of $\rho$, $K^*$ and $\phi$,
which are shown in the right panel
of Fig.~\ref{fig.mrho}. This  explains in part the very similar results between
our current value for $e_m^V$ and that in Ref.~\cite{Guo:2009hi}.

\begin{figure}
\begin{center}
 \includegraphics[angle=0,width=1.0\textwidth]{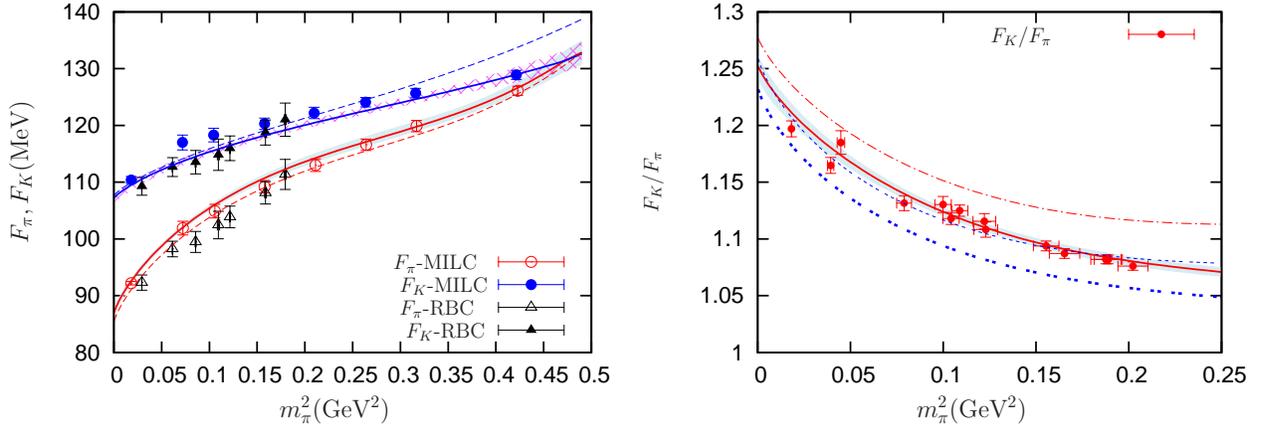}
 \caption{{\small Fit results for $F_\pi$, $F_K$,
 together with the lattice data from MILC~\cite{Davies:2003fw,Davies:2003ik},
 RBC and UKQCD~\cite{Aoki:2010dy,Arthur:2012opa} collaborations are shown in the left panel.
   The light blue and crisscross  band provide the 68\% CL regions in both plots.
In the right panel, we plot the  {  ratio  }  $F_K/F_\pi$  together
with the lattice data from Ref.~\cite{Durr:2010hr}.
The leftmost data points in each panel correspond to the physical {mass}
values, which are taken from PDG~\cite{Beringer:1900zz}.
  The solid lines in the left panel  are the fit results with physical  strange quark mass.
The solid red (lower) and double-dashed blue (lower) lines in the right-hand  plot  refer
to the expanded (Eq.~\eqref{eq.expanded-ratio}) and unexpanded  (Eq.~\eqref{fphifinal})
expressions for $F_K/F_\pi$  with $\alpha_{m_s}=1$,  respectively.
The dashed lines in the left panel correspond to $\alpha_{m_s}=1.2$.
The same applies to the  dash-dotted red (upper)
and   dashed blue (upper) lines in the right-hand  plot, which refer
to the expanded (Eq.~\eqref{eq.expanded-ratio}) and unexpanded  (Eq.~\eqref{fphifinal})
expressions for $F_K/F_\pi$,  respectively.
}}\label{fig.fpifk}
\end{center}
\end{figure}

 The light-blue and crisscross shaded areas
surrounding the solid lines in Figs.~\ref{fig.mrho} and~\ref{fig.fpifk}
represent our estimates  of the 68\% confidence level (CL)  error bands.
In order to obtain these uncertainty regions
we first generate large sets of parameter configurations by varying all our 6 fit parameters
around their central values randomly via a Monte Carlo (MC) generator;
then we use these large {amount} of parameter
configurations to calculate the $\chi^2$ and keep only the configurations with $\chi^2$ smaller than
 $\chi_0^2+\Delta \chi^2$, being $\chi_0^2$ the minimum chi-square obtained from the fit.
The 68\% CL region is given by $\Delta\chi^2=7.04$ for a 6--parameter
fit~\footnote{
 The number $\Delta\chi^2=7.04$ is obtained from
the standard multi-variable Gaussian distribution analysis
for    a     $68\%$~CL region in a 6--parameter fit~\cite{Beringer:1900zz}.
For a general $(1-\alpha)$ CL  and number of parameters $m$,
$\Delta\chi^2$ is given by
$ \alpha=\Gamma\left(\frac{m}{2},\frac{\Delta\chi^2}{2}\right)/\Gamma\left(\frac{m}{2}\right)$,
with $\Gamma(b,x)$ and $\Gamma(b)$ the incomplete gamma and Euler gamma functions,
respectively.  }.
The successful parameter configurations  provide the 68\% CL error bands.
In such a way, the correlations between the different fit parameters in Eq.~\eqref{fitresult}
have been taken into account when plotting the error bands in Figs.~\ref{fig.mrho} and \ref{fig.fpifk}.

Both our fit results and the lattice simulation data for $F_\pi$ and $F_K$
with varying pion masses are shown in the left panel of Fig.~\ref{fig.fpifk}.
The lattice data for $F_\pi$ and $F_K$ are taken from MILC~\cite{Davies:2003fw,Davies:2003ik},
RBC and UKQCD~\cite{Aoki:2010dy,Arthur:2012opa}.
Concerning the data from Refs.~\cite{Aoki:2010dy,Arthur:2012opa},
we only consider those that are simulated with the physical
strange quark mass and the unitary points.
In the right panel, we give the plots for the {  ratio }
$F_K/F_\pi$~\cite{Durr:2010hr}.
Even though the fit is performed with $m_s=m_{s,{\rm phys}}$,
we have also plotted in Fig.~\ref{fig.fpifk}
the predictions for $F_\pi$ and $F_K$ with $m_s=1.2\, m_{s,{\rm phys}}$   .
For this we have used the fit values from  Eq.~\eqref{fitresult}.
In the left panel of Fig.~\eqref{fig.fpifk},
one can see how the results with physical strange quark mass (solid lines) vary when one instead uses
$\alpha_{m_s}=1.2$ in Eq.~\eqref{mkmpi} (dashed lines).
{  In   }   the right panel of Fig.~\ref{fig.fpifk}, the  solid red line (lower) corresponds to the fit result with the
perturbative expansion of $F_K/F_\pi$ up to one loop order in Eq.~\eqref{eq.expanded-ratio}
with $m_s=m_{s,{\rm phys}}$, whereas the dash-dotted red (upper) line
uses Eq.~\eqref{eq.expanded-ratio} with $m_s=1.2\, m_{s,{\rm phys}}$.
The  blue double-dashed (lower) line represents the unexpanded value of  $F_K/F_\pi$
extracted directly from $F_\pi$ and $F_K$ from Eq.~\eqref{fphifinal} with $m_s=m_{s,{\rm phys}}$,
while  the blue dashed (upper) line uses the same unexpanded expression
but with $m_s=1.2\, m_{s,{\rm phys}}$.


  Using a value of the  strange quark mass 20\% larger  than  the physical one
only induces slight changes for $F_\pi$ and $F_K$ in the region of $m_\pi \leq 500$~MeV,
indicating the  smaller sensitivity  of these two quantities
to the linear quark mass dependence for $m_\phi^2$
assumed in Eqs.~\eqref{eq.mpi} and \eqref{mkmpi}.
Notice that  $F_\pi$ decreases when $m_s$ increases, while $F_K$ grows.
The reason is the different  way how $m_s$ enters in these two observables:
through loops and $1/N_C$ suppressed in $F_\pi$; in the valence quarks and contributing at
LO in $1/N_C$ for $F_K$.
This explains the larger shift  observed in the $F_K/F_\pi$ ratio  when varying
the strange quark mass (see the right panel in Fig.~\ref{fig.fpifk}).

\subsection{  Anatomy of the fit parameters:  correlations    }

For the scalar resonance parameters $c_d$ and $c_m$, our current
results are quite compatible with those determined in many other
processes~\cite{rcht89,Guo:2009hi,Guo:2011pa,Guo:2012yt,theta-Nc,Jamin:2001zq}.
However, the present determination of $G_V$ in Eq.~\eqref{cdgv},
is clearly lower than the usual results from phenomenological analyses,
which prefer values around 60~MeV~\cite{rcht89,Guo:2009hi,Guo:2011pa,Guo:2012yt}.
One way out of this problem is to free $G_V$ in our fit, instead of
imposing its large--$N_C$ high energy constraint from Eq.~\eqref{hegv}.

\begin{figure}[!t]
 \begin{center}
  \includegraphics[angle=0.0,width=0.4\textwidth]{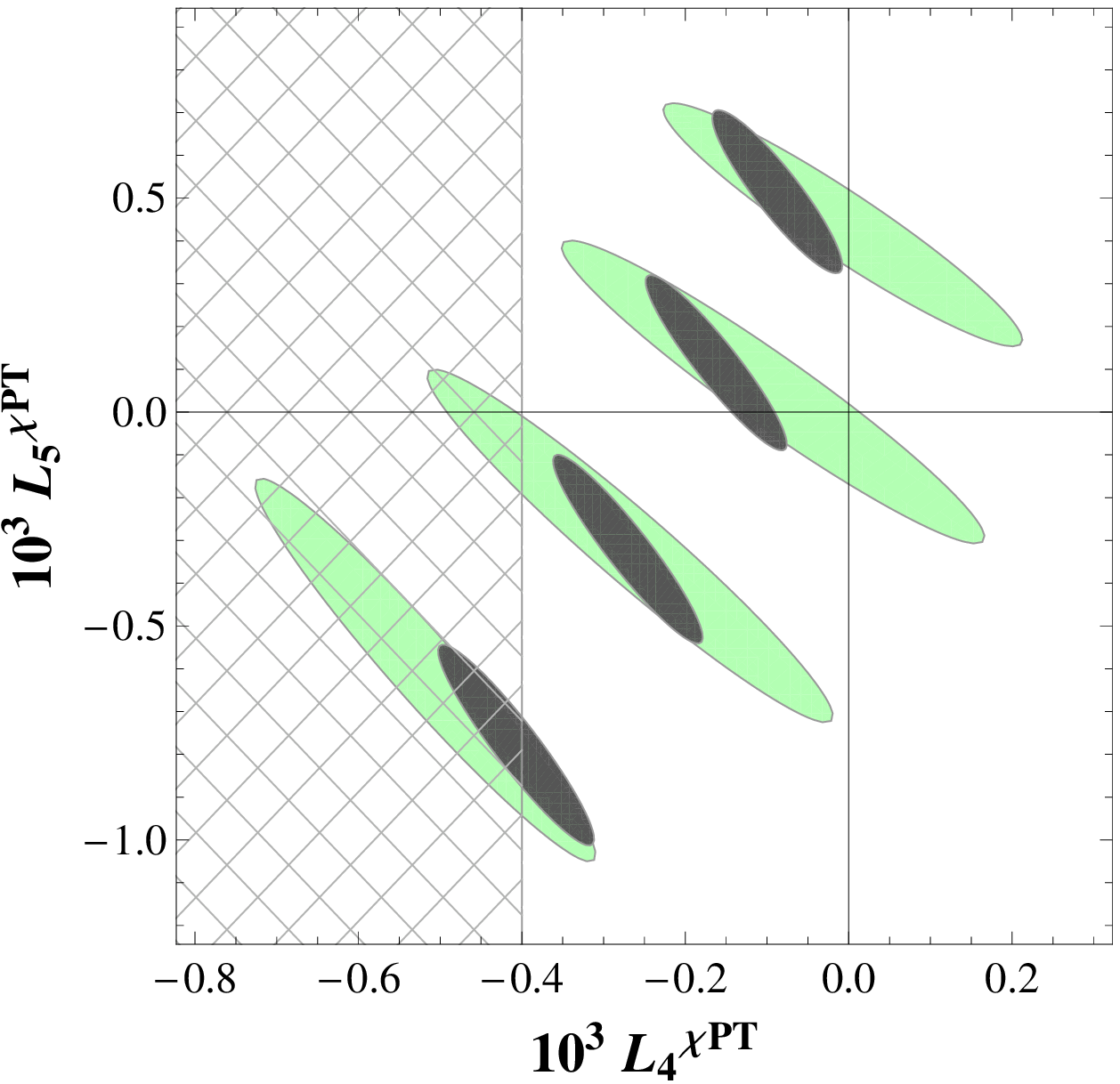}
  \caption{{\small 68\% CL regions for $L_4^{\chi PT}(\mu)$ and $L_5^{\chi PT}(\mu)$
  at $\mu=770$~MeV for the fits with $G_V$ fixed to $40,\, 50,\, 60$
  and $70$~MeV (respectively from top-right to bottom-left).
  Here and in the following plots the crisscross area  (left-hand side of this figure)
  represents the region forbidden by the paramagnetic
  inequality    $F^{n_f=3}\, <\, F^{n_f=2}$,  which implies
  $L_4^{\chi PT}(\mu)> -0.4\cdot 10^{-3}$ for
  $\mu=770$~MeV~\cite{Ecker:2013pba,DescotesGenon:1999uh}.
  The light-green ellipses come from fits to data points with $m_\pi<500$~MeV
  and the darker gray ones from the fits to all data.
  }}
  \label{fig.L4-L5-correlation}
 \end{center}
\end{figure}

   A first test is provided by setting $G_V$ to particular values.
In Fig.~\ref{fig.L4-L5-correlation},  we plot the 68\% CL regions
for $L_4^{\chi PT}$ and $L_5^{\chi PT}$ for  the fits with $G_V$
fixed to  $G_V=40, 50, 60, 70$~MeV  (ellipses from top-right to bottom-left
in Fig.~\ref{fig.L4-L5-correlation}, respectively;
Gaussianity is assumed).
This  shows how the $\rho-\pi\pi$  coupling  affects the determinations of $L_4^{\chi PT}$
and $L_5^{\chi PT}$: smaller  values of $G_V$  lead to a closer agreement with
the standard $\chi$PT phenomenology~\cite{Bijnens:2011tb,gl845}.
On the other hand, larger values of $G_V$ tend to decrease the values of both LECs;
eventually, for a large enough $\rho-\pi\pi$ coupling,
$L_5^{\chi PT}$ turns  negative and $L_4^{\chi PT}$ violates
the paramagnetic  inequality    $F^{n_f=3}\, <\, F^{n_f=2}$
($L_4^{\chi PT}(\mu)> -0.4\cdot 10^{-3}$ for $\mu=770$~MeV~\cite{Ecker:2013pba,DescotesGenon:1999uh}).
This effect cannot be attributed to an  {   inappropriate   }  description  of the kaon and pion masses
in Eqs.~\eqref{eq.mpi} and~\eqref{mkmpi} nor the fact of neglecting operators of the Lagrangian
whose  contributions to $F_\phi$  are suppressed by both $1/N_C$ and $m_\phi^4/M_S^4$.
This can be neatly observed in Fig.~\ref{fig.L4-L5-correlation}, where the black ellipses
are given by the fit to the full set of lattice data
whereas only the data with $m_\pi<500$~MeV are used in the fits that provide
the light-green regions.  Reducing the number of data points in the large pion mass region obviously leads to a
consistent  enlargement of the uncertainty regions but does not modify at all the
strong correlation  with $G_V$.

{ A second test consists on exploring two alternative versions of the high energy constraints for
the $\rho-\pi\pi$ coupling in Eq.~\eqref{hegv}:    
$G_V = F_0/\sqrt{2}$~\cite{ksrf} and $G_V = F_0/\sqrt{3}$~\cite{Guo:2007ff,Guo:2011pa}.
The former constraint corresponds to the original Kawarabayashi-Suzuki-Riazuddin-Fayyazuddin (KSRF) relation
while the latter is the extended KSRF
relation obtained by including the crossed-channel contributions
and ignoring the scalar resonances in $\pi\pi$ scattering.
We obtain $G_V\sim 58$~MeV for $G_V = F_0/\sqrt{2}$
and  $G_V\sim 47$~MeV for $G_V = F_0/\sqrt{3}$, with the chiral coupling  $F_0$ remaining  always stable and with
a value around 82~MeV.
In both situations, we confirm the findings we obtained previously when $G_V$ was fixed at the specific values
$40$, $50$, $60$ and $70$~MeV (see Fig.~\ref{fig.L4-L5-correlation}):
we observe strong anti-correlations between $L_4^{\chi PT}$ and $L_5^{\chi PT}$
and their values are strongly affected by $G_V$ in the way discussed before.
The values of  $L_4^{\chi PT}$ and $L_5^{\chi PT}$ follow closely the trend shown in Fig.~\ref{fig.L4-L5-correlation}:
the smaller $G_V$ becomes, the more negative  $L_4^{\chi PT}$ and $L_5^{\chi PT}$ turn.
Hence we conclude that our second test based on using different high
energy constraints for $G_V$ confirms our former findings and do not reveal new information with respect to
the first test, where $G_V$ was fixed at specific values.    }

  We will proceed now with our third test:   $G_V$   will be set  free and  fitted
together with the other six parameter from the previous analysis.
Statistically speaking, we do not find any significant improvement
of the fit quality by releasing this additional free parameter,
but we do see obvious changes with respect to the values in Eq.~\eqref{fitresult},
which now turn out to be
\begin{eqnarray}\label{fitresultgv}
&&F_0=83.5\pm1.5~{\rm MeV}\,, \qquad
L_4^{\chi PT}=(-0.31\pm0.10)\times 10^{-3}~\,, \qquad
L_5^{\chi PT}=(-0.46\pm0.33)\times 10^{-3}\,,
\nonumber \\
&&c_m=64.1\pm3.7~{\rm MeV}\,,\qquad
e_m^V=-0.236\pm0.004\,,  \qquad
e_m^S=-0.540\pm0.088\,,
\nonumber \\
&&G_V= 63.0\pm6.4~{\rm MeV}\,,
\end{eqnarray}
with $\chi^2/$(d.o.f)$=80.0/(52-7)$. The fit quality resulting in this case is quite
similar { to  that in  } the previous section.
   By substituting the results from  Eq.~\eqref{fitresultgv} in
the scalar form-factor high energy constraints from Eq.~\eqref{hecd} one obtains
\begin{eqnarray}\label{cd-fit2}
c_d &=& \, (\, 27.2 \, \pm \,  1.8\, )\, {\rm MeV}\,.
\end{eqnarray}

\begin{figure}
 \begin{center}
  \includegraphics[angle=0.0,width=0.9\textwidth]{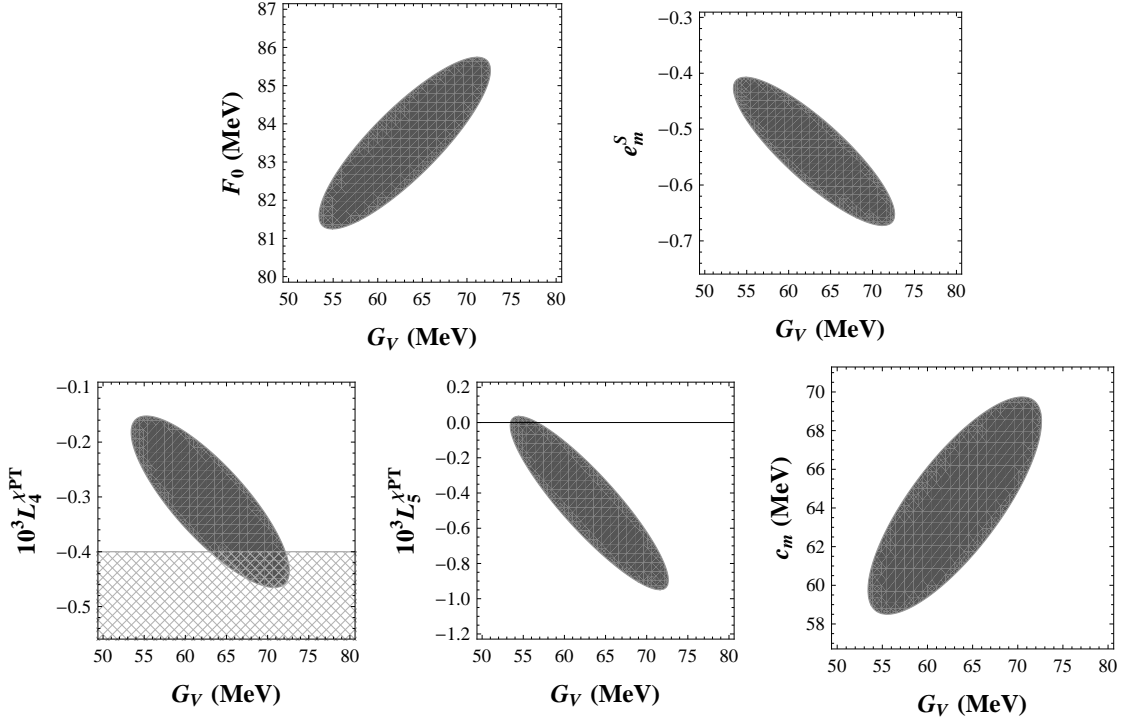}
  \caption{{\small
     $68\%$~CL regions for $G_V$ and  other parameter.
        All these plots  correspond to the 7-parameter analysis in Eq.~\eqref{fitresultgv},
        where $G_V$ is also fitted.
  } }\label{fig.cor1}
 \end{center}
\end{figure}

\begin{figure}
 \begin{center}
  \includegraphics[angle=0.0,width=0.65\textwidth]{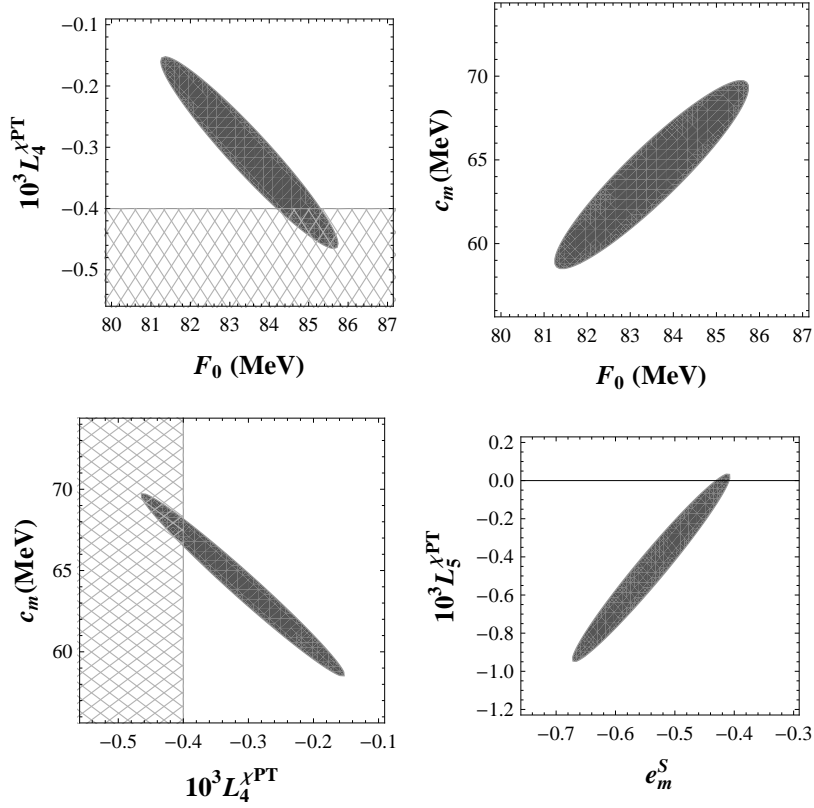}
  \caption{{\small
     $68\%$~CL regions  for sets of two couplings.
        All these plots  correspond to the 7-parameter analysis in Eq.~\eqref{fitresultgv},
        where $G_V$ is also fitted.
}}\label{fig.cor2}
 \end{center}
\end{figure}

The {most striking}   change happens for $L_5^{\chi PT}$, whose sign becomes negative.
However, according to most  phenomenological determinations of $L_5^{\chi PT}$
in literature~\cite{Bijnens:2011tb,Ecker:2010nc,Ecker:2013pba}
its value must be positive.
Also R$\chi$T predicts a positive $L_5^{\chi PT}$ at large $N_C$~\cite{rcht89}.
Hence the resulting parameters in Eq.~\eqref{fitresultgv} do not seem to
correspond to the physical solution. The reason behind this
is the strong correlations between different parameters:
we observe that the parameter $G_V$ is strongly
correlated with all of the other parameters.
 The only exception is $e_m^V$, which is mostly uncorrelated and is essentially
determined by the $\rho(770)-K^*(892)-\phi(1020)$ splitting.
The correlations are summarized in Figs.~\ref{fig.cor1}
and \ref{fig.cor2} (Gaussianity is assumed).
In Fig.~\ref{fig.cor1} we provide the correlation between $G_V$ and the
other fit variables. One can  clearly
see that the parameter $G_V$, which rules  the $\rho - \pi\pi$
interaction vertex in the chiral limit,
is highly correlated  with almost
{  all the other parameters.   }By observing this plots, one can easily understand
why we have obtained
{  such  different  values for  }
$L_5^{\chi PT}$ in Eqs.~\eqref{fitresult}
{(}with $G_V$ constrained through Eq.~\eqref{hegv}{)}
and \eqref{fitresultgv} (free $G_V$):
the  values for $G_V$  are very different in the two fits
and a positive (negative) $L_5^{\chi PT}$ requires a small (large) value for $G_V$
(see bottom-center panel in Fig.~\ref{fig.cor1}).

In the top left panel in Fig.~\ref{fig.cor2},  one can clearly observe
an  evident anti-correlation between  $F_0$ and $L_4^{\chi PT}$,
noticed in previous works~\cite{Ecker:2010nc,Ecker:2013pba}.
In addition, we observe a strong anti-correlation for $L_4^{\chi PT}-c_m$
and an obvious correlation for $L_5^{\chi PT}-e_m^S$, as shown in the two panels
in the bottom row of Fig.~\ref{fig.cor2}.
In Ref.~\cite{DescotesGenon:1999uh}, a lower bound on the value of $L_4^{\chi PT}$
has been proposed by requiring that the pNGB
decay constant in $SU(3)$ chiral limit must be  smaller than the decay  constant
in the  $SU(2)$ limit.   This gives  the inequality $L_4^{\chi PT}>-0.4\times10^{-3}$
for $\mu=770$~MeV~\cite{Ecker:2013pba}.
It is interesting to point out that this lower bound from $L_4^{\chi PT}$
leads to lower or upper bounds for  some of the parameters considered in our work
because of the strong correlations.
This can be roughly read from Figs.~\ref{fig.cor1} and \ref{fig.cor2}:
$G_V<72$~MeV, $F_0<86$~MeV and $c_m<68$~MeV.
On the other hand, one can observe in Figs.~\ref{fig.cor1} and~\ref{fig.cor2}
that in order to have  a positive  $L_5^{\chi PT}$ one has the rough bounds
$G_V<60$~MeV and $e_m^S>-0.45$.
Combining the paramagnetic inequality for
$L_4^{\chi PT}$~\cite{DescotesGenon:1999uh,Ecker:2013pba}
and the {  phenomenological   }  bound $L_5^{\chi PT}>0$
leads to the rough estimates $F_0<86$~MeV, $G_V<60$~MeV, $c_m<68$~MeV and $e_m^S>-0.45$.

{     In order to further test the relations between $G_V$ and other parameters,
we want to see the impact of including a second scalar nonet.     }
The contributions from the second scalar nonet to the decay constants
in Appendix~\ref{app.diagrams} and the matching conditions in Eqs.~\eqref{eq.Ftilde-matching}-\eqref{l5chptmatching} share
the same expressions as the lowest scalar multiplet, but with obvious replacements of the couplings $c_d$, $c_m$,
$e_m^S$ by $c_d'$, $c_m'$ and $e_m^{S'}$. The chiral limit resonance mass $M_{S'}$ of the excited nonet $S'$
should be replaced as well. The introduction of the second scalar nonet will also affect the high energy constraints
in Eqs.~\eqref{hecd} and \eqref{hegv}, which now become~\cite{Jamin:2001zq,Guo:2007ff}
\begin{eqnarray}\label{hecds1}
c_d &=&   {   \frac{F_0^2-4c_d'c_m'}{4c_m}\,,   }    \\
G_V &=&    {   \sqrt{\frac{F_0^2-2c_d^2-2{c_d'}^2}{3}}\,.   }    \label{hegvs1}
\end{eqnarray}
{    Phenomenologically, the $S'$ parameters are poorly known in the literature
and we do not expect to obtain precise values from our analysis.
In order to perform our quantitative estimate of the role of the
second scalar nonet, we take the part of the outcomes  from Ref.~\cite{Jamin:2000wn} as inputs.
More precisely, we take  $c_m'=c_d'$ and $M_{S'}=2.57$~GeV
(preferred fit values from Ref.~\cite{Jamin:2000wn}, Eq.(6.10) therein).~\footnote{
  {    We point out that the constraints $c_m=c_d$ and $c_m'=c_d'$ in Ref.~\cite{Jamin:2001zq} are obtained by considering the linear quark mass corrections in
the minimal R$\chi$T framework (only operators with one resonance field).
These two constraints do not hold any more if general R$\chi$T  operators with any number of resonance
fields~\cite{Op6-RChT} are included in the Lagrangian.
This is the reason why  we do not impose the constraint $c_m=c_d$ in our previous discussion
with  only the lightest scalar nonet.
We will nevertheless employ the relation $c_m'=c_d'$ in our numerical estimate in order to  stabilize the fit
with two scalar nonets and to get a general idea of the impact of the second scalar multiplet.  }     }
The mass splitting parameter $e_m^{S'}$  is even worse known than $c_d'$ and $c_m'$. In our rough analysis
we will set its value to zero.
Thus, we only have one free parameter $c_d'$ from the second scalar nonet. In Refs.~\cite{Jamin:2000wn,Jamin:2001zq},
$c_d'$ was obtained through the constraint $F_0^2=4(c_dc_m+c_d'c_m')$ by using $F_0=F_\pi=92.4$~MeV,
whereas in our work $F_0$ is truly the pion decay constant in the chiral limit.
Hence, instead of taking the result of $c_d'$ from Ref.~\cite{Jamin:2000wn},
we fit its value together with the chiral coupling $F_0$.
}

{  The fit result with the new constraints in Eqs.~\eqref{hecds1} and \eqref{hegvs1} (two scalar multiplets)
turn out to be quite similar to the outcomes in Eq.~\eqref{fitresult}
with only one scalar nonet in the high-energy constraints~\eqref{hecd} and~\eqref{hegv}.
The additional coupling becomes $c_d'\simeq (11\pm 30)$~MeV, a value compatible with
the preferred determination in Ref.~\cite{Jamin:2000wn} and alternative fits therein.
In this work we reconfirm the large uncertainty for $c_d'$ obtained  from $K\pi$ scattering~\cite{Jamin:2000wn}.
We have also tried other fits where high-energy relation~\eqref{hegvs1} is released and $G_V$ is freed and fitted.
The $\rho-\pi\pi$ coupling has been also set in later fits to the particular values
$G_V = F_0/\sqrt{2}$~\cite{ksrf} and $G_V = F_0/\sqrt{3}$~\cite{Guo:2007ff,Guo:2011pa}.
In all these cases the results tend to produce small central values for $c_d'$ but with large uncertainties.
As a result, the inclusion of the second scalar nonet barely changes
our conclusions derived previously with only one scalar nonet.    }

In summary:  the present determination for $F_0$,
the $n_f=3$ pNGB decay constant in the chiral limit, is rather  stable,
{ ranging from  78 to 86~MeV for any value of  $G_V$ in the range $40\sim 70$~MeV.   }
Our current determinations of the $\chi$PT LECs $L_4^{\chi PT}$ and $L_5^{\chi PT}$ can not be pinned down to a precise range due to
their strong correlations with the resonance couplings, which are typically determined through some phenomenological processes with non-negligible uncertainties.
Among the various resonance couplings, $G_V$ turns out to be the crucial one to prevent us from making
  precise determinations.
In the case of imposing the high energy constraint on $G_V$ from Eq.~\eqref{hegv},
obtained from  the discussion of the partial wave $\pi\pi$ scattering
at LO in $1/N_C$~\cite{Guo:2007ff},
the corresponding fit results
in Eq.~\eqref{fitresult} are more or less compatible with the state-of-art
determinations of the $\chi$PT LECs.
{We  } regard these results
as our preferred ones in this work.
Nonetheless, one should always bear in mind the strong correlations shown
in Figs.~\ref{fig.cor1} and \ref{fig.cor2}.

\section{Conclusions}
\label{sec.conclusions}

The aim of this work is to provide a first quantitative test  of
the potentiality of these type of hadronic observables, such as the pNGB decay constants,
for the study of resonance properties.
We have calculated the pion and kaon weak decay constants
within the framework of R$\chi$T up to NLO in $1/N_C$, this is, up to the one-loop level.
In addition to the octet of light pNGB, we have explicitly included
the singlet $\eta_1$  and the lightest vector and scalar resonance multiplets surviving at large $N_C$.
However, we want to remark that the errors provided here should be considered with
quite some care, as we have combined data from various simulation groups,
{   ignoring correlations and  systematic  and lattice spacing uncertainties.  }

Our one-loop expressions for $F_\pi$ and $F_K$ in R$\chi$T  have been properly matched  to
$SU(3)$ $\chi$PT up to  $\cO{(p^4)}$ in the small quark mass regime,
providing  prediction for the chiral LECs in terms of the R$\chi$T parameters.
As higher order corrections from $\chi$PT are partly incorporated
through the resonance loops,  the present calculation
provides an alternative approach which complements  previous $\chi$PT
analyses~\cite{Ecker:2013pba,FP-Op6,lattice-TBC,Bernard:2011}.
The price to pay in the latter is, however,  the vast amount of $\chi$PT couplings one needs to
consider in the full $\cO(p^6)$ expression.
In our work, the resonances are assumed to play a crucial role instead,
ruling the dynamics of the decay constant.

We have extended the work from Ref.~\cite{Soto:2011ap} (which incorporated the scalar
effects at one loop)  by considering also the impact of vector resonances in the loops.
One of the fundamental  conclusions in our study is that the vectors play a crucial role in the
one-loop decay constant, being crucial parameters such as $F_0$, $L_4^{\chi PT}$  and $L_5^{\chi PT}$
very correlated with the value of the $\rho-\pi\pi$ coupling $G_V$.
Low  values of $G_V$, around 40~MeV, lead to larger values of $L_4^{\chi PT}$ and $L_5^{\chi PT}$,
in closer agreement with standard $\chi$PT determinations~\cite{Bijnens:2011tb}.
Due to the $L_4^{\chi PT}\leftrightarrow F_0$ anti-correlation this yields a small value for $F_0$,
around 80~MeV.
On the other hand, a  $G_V$ coupling in the range $60\sim70$~MeV seems to be in better agreement
with   vector resonance phenomenology~\cite{rcht89,SD-RChT,Guo:2007ff,Guo:2009hi}
but generates  a far too negative value for both $L_4^{\chi PT}$ and $L_5^{\chi PT}$,
in clear contradiction with $\chi$PT determinations~\cite{Bijnens:2011tb}
and QCD paramagnetic inequalities ~\cite{DescotesGenon:1999uh}
($L_4^{\chi PT}>-0.4\cdot 10^{-3}$
for $\mu=770$~MeV~\cite{Ecker:2013pba}). Nonetheless,
in spite of this big effect on the $\cO(p^4)$ LECs, $F_0$ happens to be very stable
and only rises up to roughly 85~MeV.
Clearly, this interplay between vector resonance loops and $\chi$PT loops
deserves further investigation in future works.

In the fit where $G_V$ is fixed to 40, 50, 60 and 70~MeV  we observe clearly
how   the coupling $F_0$ evolves from $80$ up to $85$~MeV.
Although the  upper value  is compatible with recent $\cO(p^6)$ estimates~\cite{Ecker:2013pba},
other analyses favor values of $F_0$ below 80~MeV~\cite{Bernard:2011}.
In general, there is no agreement yet (see FLAG's review~\cite{FLAG:2013} and references therein)
and the strong anti-correlation between $F_0$ and $L_4^{\chi PT}$ found here and in previous
works~\cite{Bijnens:2011tb,Ecker:2013pba} transfers this uncertainty
to the $\cO(p^4)$ LEC $L_4^{\chi PT}$.


The analysis of
$F_\pi$, $F_K$~\cite{Davies:2003fw,Davies:2003ik,Aoki:2010dy,Arthur:2012opa}
and $F_K/F_\pi$~\cite{Durr:2010hr}
was carried out in combination with a study of the quark mass dependence of
the $\rho(770)$, $\eta$ and $\eta'$ masses.
The simple quark mass dependence of the vector multiplet mass introduced through $e_m^V$
perfectly accommodates the $M_\rho$ lattice  data~\cite{Lang:2011mn,Aoki:2011yj,Pelissier:2012pi,Dudek:2012xn}
and the observed splitting  of the physical vector multiplet~\cite{Beringer:1900zz}
(Fig.~\ref{fig.mrho}).  Likewise, the LO prediction for  the $\eta-\eta'$ mixing is found to be
in  reasonable agreement
with lattice data~\cite{Michael:2013vba,Michael:2013gka,Christ:2010dd,Dudek:2011tt,Gregory:2011sg}
(Fig.~\ref{fig.meta}).
We find that our theoretical formulas can   reproduce the lattice data from the physical pion mass
up to roughly  $m_\pi=700$~MeV.   This result gives support to the linear relation between
the pNGB and quark masses from Eqs.~\eqref{eq.mpi} and~\eqref{mkmpi}, assumed all along the article.


Based on the promising fact that the present framework performs a  reasonable chiral extrapolation
for $F_\pi$ and $F_K$ within a broad range of
pion masses, a similar study on the masses of $\pi$, $K$, or even $\eta$ and $\eta'$
should be pursued
{ within R$\chi$T up to NLO in $1/N_C$.     
This would  also allow us to  go beyond the  linear quark mass dependence
considered  for  the squared masses  of the pion and kaon in this article.    }
We think this might help to set further and more stringent constraints
on the  low energy constants of the $\chi$PT Lagrangian.

\section*{Acknowledgements}
We would like to thank   Alberto Ramos and
Pere Masjuan for useful discussions,
specially on the detailed explanations of the lattice simulation data.
This work is partially funded by the grants
National Natural Science Foundation of China (NSFC) under contract No. 11105038,
Natural Science Foundation of Hebei Province with contract No. A2011205093,
Doctor Foundation of Hebei Normal University with contract No. L2010B04,
  the Spanish Government  and ERDF funds from the European Commission
[FPA2010-17747,      
SEV-2012-0249,
CSD2007-00042] and the Comunidad de Madrid [HEPHACOS    S2009/ESP-1473].

\appendix

\section{Feynman integrals}
\label{app.Feynman-int}

The explicit expressions for the loop functions used in this work are given by
\begin{eqnarray}
A_0(m^2)&=&
 \Int \Frac{d^dk}{(2\pi)^d} \, \Frac{1}{k^2-m^2}
\quad =\quad \frac{i}{16\pi^2} m^2\bigg( \frac{1}{\hat\epsilon}  - \ln{\frac{m^2}{\mu^2}} \bigg)\,,
\nn\\
\nonumber \\
B_0(q^2,M^2,m^2)&=&
 \Int \Frac{d^dk}{(2\pi)^d} \, \Frac{1}{(k^2-m^2)\, ((q-k)^2-M^2)}
\nn\\
&&\quad=\quad
\frac{i}{16\pi^2}\bigg[ \frac{1}{\hat\epsilon} + 1 - \frac{1}{2}\ln{\frac{M^2 m^2}{\mu^4}}
+  \frac{\Delta}{2s}\ln{\frac{ m^2}{M^2}} -\frac{\nu}{2s} \ln\frac{(\Sigma-s-\nu)^2}{4M^2 m^2} \bigg]\,,
\nn\\
\nonumber \\
B_0'(s,M^2,m^2)&=&\frac{d B_0(s,M,m)}{d s}= \frac{i}{16\pi^2}\bigg[
 -\frac{\Delta}{2s^2}\ln{\frac{ m^2}{M^2}} -\frac{1}{s} + \frac{\Delta^2-\Sigma s}{2\nu s^2}\ln\frac{(\Sigma-s-\nu)^2}{4M^2 m^2} \bigg]\,,
\nonumber \\
\end{eqnarray}
where
\begin{eqnarray}
\frac{1}{\hat\epsilon}&=&
 \mu^{-2\epsilon}  \, \left( \frac{1}{\epsilon}-\gamma_E + \ln{4\pi}+1
\right)  \quad=\quad \Frac{1}{\epsilon} -\gamma_E + \ln{4\pi}+1 - \ln\mu^2
+\cO(\epsilon) \, ,
 \qquad (\epsilon= 2-\frac{D}{2})\,,
\nonumber \\
\Delta&=& M^2-m^2\,,\qquad \Sigma=M^2+m^2\,, \qquad \nu=\sqrt{\big[ s- (M+m)^2 \big] \big[ s- (M-m)^2 \big]}\,.
\end{eqnarray}

\section{$\eta_1-\eta_8$ mixing}
\label{app.eta-etap}

After the diagonalization of $\eta_1-\eta_8$ at leading order, we have the physical $\eta$ and
$\eta'$ states at this order and their masses and the mixing angle can be found in many references in literature, such as Ref.~\cite{Guo:2011pa}.
We give the explicit formulas for the sake of completeness
\begin{eqnarray}
m_{\eta}^2 &=& \frac{M_0^2}{2} + m_K^2
- \frac{\sqrt{M_0^4 - \frac{4 M_0^2 \Delta^2}{3}+ 4 \Delta^4 }}{2} \,, \label{defmetab2}  \\
m_{\eta'}^2 &=& \frac{M_0^2}{2} + m_K^2
+ \frac{\sqrt{M_0^4 - \frac{4 M_0^2 \Delta^2}{3}+ 4 \Delta^4 }}{2} \,, \label{defmetaPb2}  \\
\sin{\theta} &=& -\left( \sqrt{1 +
\frac{ \big(3M_0^2 - 2\Delta^2 +\sqrt{9M_0^4-12 M_0^2 \Delta^2 +36 \Delta^4 } \big)^2}{32 \Delta^4} } ~\right )^{-1}\,,
\label{deftheta0}
\end{eqnarray}
with $\Delta^2 = m_K^2 - m_\pi^2$.
  Notice that $m_\eta$, $m_{\eta'}$ and $\theta$ are
fully determined at this order by $m_\pi$, $m_K$ and $M_0$.

   In the ideal mixing case ($M_0=0$) one gets
$m_\eta^2 =m_\pi^2$, $m_{\eta'}^2 = 2 m_K^2-m_\pi^2$
and ${  \sin\theta=-\sqrt{2/3}  }$.
On the other hand, in the chiral limit $m_\pi,\, m_K \to 0$ the physical masses and mixing become
$m_\eta^2=0$, $m_{\eta'}^2=M_0^2$ and $\theta=0$.

   \section{Feynman diagrams up to NLO in $1/N_C$}
\label{app.diagrams}

\subsection{The pion self-energy}
\label{app.pi-self}

As shown in Fig.~\ref{fig.se}, there are three types of Feynman diagrams contributing to the pNGB self-energy $\Sigma^{\pi}$.
For the diagram (a)  in this figure, the explicit calculation from Lagrangian in Eq.~\eqref{lagchpt4} leads to
\begin{eqnarray}
\Sigma^{\pi-a}=&&
 \bigg(1- \Frac{\widetilde{F}^2}{F_0^2} \bigg)p^2
- \bigg(1-\Frac{\hat{F}^2}{F_0^2}\bigg)m_{\pi}^2 \,\,\, +\,\,\,
\frac{4\widetilde{L}_{12}}{F_0^2}(p^2-m_\pi^2)^2-\frac{8\widetilde{L}_{11}}{F_0^2}m_\pi^2(p^2-m_\pi^2)
\nonumber \\ &&
\hspace*{-0.75cm} -\frac{8\widetilde{L}_4}{F_0^2}(2m_K^2+m_\pi^2)p^2
-\frac{8}{F_0^2}(\widetilde{L}_5+\frac{c_dc_m}{M_S^2})m_\pi^2\,p^2
+\frac{16\widetilde{L}_6}{F_0^2}(2m_K^2+m_\pi^2)m_\pi^2+\frac{16}{F_0^2}(\widetilde{L}_8+\frac{c_m^2}{2M_S^2})m_\pi^4\,.
\nonumber \\
\end{eqnarray}
  where we have used the linear relations~\eqref{eq.mpi} and~\eqref{mkmpi}
to rewrite the quark masses in terms of the pion and kaon masses.
The tree-level contribution from the operators
in    the second  line of Eq.~\eqref{lagsalarshift} have been also taken into account.

About the diagram (b) in Fig.~\ref{fig.se}, its contribution to the pion self-energy is the same as in $U(3)$ $\chi$PT, which
is calculated by using leading order Lagrangian in Eq.~\eqref{lolagrangian} and reads
\begin{eqnarray}
\Sigma^{\pi-b} &&= \frac{i}{3F_0^2}(2p^2-\frac{m_\pi^2}{2})A_0(m_\pi^2)
+ \frac{i}{3F_0^2}(p^2-m_\pi^2)A_0(m_K^2) + \frac{-i}{2F_0^2}\frac{(c_\theta-\sqrt2s_\theta)^2}{3}m_\pi^2A_0(m_\eta^2)
\nonumber \\ &&
+ \frac{-i}{2F_0^2}\frac{(\sqrt2c_\theta+s_\theta)^2}{3}m_\pi^2A_0(m_{\eta'}^2)\,.
\end{eqnarray}

{ The diagram (c) in Fig.~\ref{fig.se}  receives   }   contributions both from scalar and vector resonances.
Let us take the self-energy for the $\pi^-$ for illustration.
There are five possible combinations of scalar resonance and
pseudoscalar meson running inside the loop:
$\sigma\pi^-$, $\kappa^- K^0$, $\kappa^0 K^-$, $a_0^-\eta$ and
$a_0^-\eta'$, which will be labeled as $-i\Sigma^{\pi-cSj}$, with $j=1,2,3,4,5$,      {  respectively.    }
About the vector, there are four possible combinations: $\rho^-\pi^0$, $\rho^0\pi^-$, $K^{*0} K^-$ and $K^{*-} K^0$ , which
will be labeled as $-i\Sigma^{\pi-cVj}$, with $j=1,2,3,4$,      {  respectively.    }

The explicit results of $\Sigma^{\pi-cSj}$ for $j=1,2,3,4,5$ are
\begin{eqnarray}\label{zpis1f}
\Sigma^{\pi-cS1}&& =
\frac{i2 c_d^2}{F_0^4} \bigg[\, (3 p^2 + m_\pi^2- M_\sigma^2)A_0(M_\sigma^2) -(m_\pi^2+ p^2 - M_\sigma^2)A_0(m_\pi^2)
\nonumber \\&& \qquad \qquad
+(m_\pi^2+ p^2 - M_\sigma^2)^2 B_0(p^2, M_\sigma^2, m_\pi^2)   \,\bigg]
\nonumber \\&& \quad
-\frac{i8 c_d c_m}{F_0^4 M_S^2}m_\pi^2 M_\sigma^2 \bigg[\,
(m_\pi^2+ p^2 - M_\sigma^2) B_0(p^2, M_\sigma^2, m_\pi^2)+ A_0(M_\sigma^2) -A_0(m_\pi^2) \,\bigg]
\nonumber \\&& \quad
+\frac{i8 c_m^2}{F_0^4 M_S^4}m_\pi^4  \bigg[\,
M_\sigma^4 B_0(p^2, M_\sigma^2, m_\pi^2)+ (m_\pi^2+ p^2 + M_\sigma^2)A_0(m_\pi^2) \,\bigg]\,,
\end{eqnarray}
\begin{eqnarray}
\Sigma^{\pi-cS2}=\Sigma^{\pi-cS3} && =
\frac{i c_d^2}{F_0^4} \bigg[\, (3 p^2 + m_K^2- M_\kappa^2)A_0(M_\kappa^2) -(m_K^2+ p^2 - M_\kappa^2)A_0(m_K^2)
\nonumber \\&& \qquad \qquad
+(m_K^2+ p^2 - M_\kappa^2)^2 B_0(p^2, M_\kappa^2, m_K^2)   \,\bigg]
\nonumber \\&& \quad
-\frac{i2 c_d c_m}{F_0^4 M_S^2} \bigg\{\,
\big[ 2m_\pi^2 M_\kappa^2+(m_K^2-m_\pi^2)(M_\kappa^2-s-m_K^2) \big] A_0(M_\kappa^2)
\nonumber \\&&  \qquad \qquad
+\big[-2m_\pi^2 M_\kappa^2+(m_K^2-m_\pi^2)(m_K^2-s-M_\kappa^2) \big] A_0(m_K^2)
\nonumber \\&&  \qquad \qquad
+\big[2m_\pi^2 M_\kappa^2(s+m_K^2-M_\kappa^2)
\nonumber \\&&  \qquad \qquad\qquad
+(m_K^2-m_\pi^2)(s^2-M_\kappa^4-m_K^4+2m_K^2 M_\kappa^2) \big] B_0(p^2, M_\kappa^2, m_K^2)\,\bigg\}
\nonumber \\&& \quad
+\frac{i c_m^2}{F_0^4 M_S^4}  \bigg\{\,
\big[-4m_\pi^2(m_K^2-m_\pi^2)M_\kappa^2+(m_K^2-m_\pi^2)^2(m_K^2-s-M_\kappa^2) \big] A_0(M_\kappa^2)
\nonumber \\&&
+\big[4m_\pi^4 (s+m_K^2+M_\kappa^2)+4m_\pi^2(m_K^2-m_\pi^2)(2s+M_\kappa^2)
\nonumber \\&& \quad
+(m_K^2-m_\pi^2)^2(3s+M_\kappa^2-m_K^2)  \big] A_0(m_K^2)
\nonumber \\&&
+\big[ 4m_\pi^4 M_\kappa^4+4m_\pi^2(m_K^2-m_\pi^2)(s+M_\kappa^2-m_K^2)M_\kappa^2
\nonumber \\&&  \quad
+(m_K^2-m_\pi^2)^2(s+M_\kappa^2-m_K^2)^2 \big] B_0(p^2, M_\kappa^2, m_K^2)
 \,\bigg\} \,,
\end{eqnarray}
\begin{eqnarray}
\Sigma^{\pi-cS4} && =
\frac{(c_\theta-\sqrt2s_\theta)^2}{3}\bigg\{
 \frac{i 2c_d^2}{F_0^4} \bigg[\, (3 p^2 + m_\eta^2- M_a^2)A_0(M_a^2) -(m_\eta^2+ p^2 - M_a^2)A_0(m_\eta^2)
\nonumber \\&& \qquad \qquad
+(m_\eta^2+ p^2 - M_a^2)^2 B_0(p^2, M_a^2, m_\eta^2)   \,\bigg]
\nonumber \\&& \quad
-\frac{i8 c_d c_m}{F_0^4 M_S^2}m_\pi^2 M_a^2 \bigg[\,
(m_\eta^2+ p^2 - M_a^2) B_0(p^2, M_a^2, m_\eta^2)+ A_0(M_a^2) -A_0(m_\eta^2) \,\bigg]
\nonumber \\&& \quad
+\frac{i8 c_m^2}{F_0^4 M_S^4}m_\pi^4  \bigg[\,
M_a^4 B_0(p^2, M_a^2, m_\eta^2)+ (m_\eta^2+ p^2 + M_a^2)A_0(m_\eta^2) \,\bigg] \,\,\bigg\}\,,
\end{eqnarray}
\begin{eqnarray}
\Sigma^{\pi-cS5} && =
\frac{(\sqrt2c_\theta+s_\theta)^2}{3}\bigg\{
 \frac{i 2c_d^2}{F_0^4} \bigg[\, (3 p^2 + m_{\eta'}^2- M_a^2)A_0(M_a^2) -(m_{\eta'}^2+ p^2 - M_a^2)A_0(m_{\eta'}^2)
\nonumber \\&& \qquad \qquad
+(m_{\eta'}^2+ p^2 - M_a^2)^2 B_0(p^2, M_a^2, m_{\eta'}^2)   \,\bigg]
\nonumber \\&& \quad
-\frac{i8 c_d c_m}{F_0^4 M_S^2}m_\pi^2 M_a^2 \bigg[\,
(m_{\eta'}^2+ p^2 - M_a^2) B_0(p^2, M_a^2, m_{\eta'}^2)+ A_0(M_a^2) -A_0(m_{\eta'}^2) \,\bigg]
\nonumber \\&& \quad
+\frac{i8 c_m^2}{F_0^4 M_S^4}m_\pi^4  \bigg[\,
M_a^4 B_0(p^2, M_a^2, m_{\eta'}^2)+ (m_{\eta'}^2+ p^2 + M_a^2)A_0(m_{\eta'}^2) \,\bigg] \,\,\bigg\}\,.
\end{eqnarray}

For the vector contributions, we have
\begin{eqnarray}
\Sigma^{\pi-cV1}=\Sigma^{\pi-cV2}&& =
\frac{i G_V^2}{F_0^4} \bigg\{\, -( p^2 - m_\pi^2+ M_\rho^2)A_0(M_\rho^2) -(p^2+m_\pi^2 - M_\rho^2)A_0(m_\pi^2)
\nonumber \\&& \qquad \qquad
+\big[( p^2-m_\pi^2 +M_\rho^2)^2-4p^2 M_\rho^2 \big] B_0(p^2, M_\rho^2, m_\pi^2)   \,\bigg\}\,,
\end{eqnarray}
\begin{eqnarray}
\Sigma^{\pi-cV3}=\Sigma^{\pi-cV4}&& =
\frac{i G_V^2}{2F_0^4} \bigg\{\, -( p^2 - m_K^2+ M_{K^*}^2)A_0(M_{K^*}^2) -(p^2+m_K^2 - M_{K^*}^2)A_0(m_K^2)
\nonumber \\&& \qquad \qquad
+\big[( p^2-m_K^2 +M_{K^*}^2)^2-4p^2 M_{K^*}^2 \big] B_0(p^2, M_{K^*}^2, m_K^2)   \,\bigg\}\,.
\end{eqnarray}

\subsection{The kaon self-energy }
\label{app.K-self}

The calculation of the kaon self-energy is similar {to}   the pion case.
The corresponding self-energy function from the type (a) diagram in Fig.~\ref{fig.se} is
\begin{eqnarray}
\Sigma^{K-a}=&&
   \bigg(1-\Frac{\widetilde{F}^2}{F_0^2}\bigg)p^2- \bigg(1-\Frac{\hat{F}^2}{F_0^2}\bigg)m_{K}^2
\,\,\, +\,\,\,
\frac{4\widetilde{L}_{12}}{F_0^2}(p^2-m_K^2)^2-\frac{8\widetilde{L}_{11}}{F_0^2}m_K^2(p^2-m_K^2)
\nonumber \\&&
\hspace*{-1.0cm}  -\frac{8\widetilde{L}_4}{F_0^2}(2m_K^2+m_\pi^2)p^2-\frac{8}{F_0^2}(\widetilde{L}_5+\frac{c_dc_m}{M_S^2})m_K^2\,p^2
+\frac{16\widetilde{L}_6}{F_0^2}(2m_K^2+m_\pi^2)m_K^2+\frac{16}{F_0^2}(\widetilde{L}_8
+\frac{c_m^2}{2M_S^2})m_K^4\,,
\nonumber \\
\end{eqnarray}
  where we have used the linear relations~\eqref{eq.mpi} and~\eqref{mkmpi}
to rewrite the quark masses in terms of the pion and kaon masses.

Again, the diagram (b) in Fig.~\ref{fig.se} leads to the same results as in $U(3)$ $\chi$PT, which is given by
\begin{eqnarray}
\Sigma^{K-b} &&= \frac{i}{2F_0^2}(p^2-m_K^2)A_0(m_K^2)
+\frac{i}{4F_0^2}(p^2-m_K^2)A_0(m_\pi^2)  \nonumber \\ &&
+ \frac{i}{12F_0^2}\bigg\{ 3c_\theta^2 (p^2+m_\eta^2)-\big[ (3c_\theta^2+4\sqrt2c_\theta s_\theta +4s_\theta^2 )m_K^2
-(c_\theta^2+2\sqrt2c_\theta s_\theta )m_\pi^2\big]  \bigg\} A_0(m_\eta^2)
\nonumber \\ &&
+  \frac{i}{6F_0^2}\bigg\{ \frac{3s_\theta^2}{2}(p^2+m_{\eta'}^2)
-\frac{1}{2}\big[ (4c_\theta^2-4\sqrt2c_\theta s_\theta +3s_\theta^2 )m_K^2
+(2\sqrt2c_\theta s_\theta -s_\theta^2)m_\pi^2 \big] \bigg\}A_0(m_{\eta'}^2)\,. \nonumber \\
\end{eqnarray}

About the diagram (c) in Fig.~\ref{fig.se}, let us take the self-energy for the $K^-$ for illustrating purpose.
There are eight possible combinations of scalar resonance and
pseudoscalar meson running inside the loop: $\sigma K^-$, $\sigma' K^-$, $a_0^0 K^-$, $a_0^-\bar{K}^0$,
$\kappa^-\pi^0$, $\kappa^0\pi^-$, $\kappa^-\eta$ and $\kappa^-\eta'$, which will be labeled as $-i\Sigma^{K-cSj}$,
with $j=1,2,3,4,5,6,7,8$, respectively.
For the vector case, there are also eight possible combinations: $\rho^0 K^-$, $\rho^-\bar{K}^0$, $\omega K^-$, $\phi K^-$,
$K^{*-}\pi^0$, $\bar{K}^{*0}\pi^-$, $K^{*-}\eta$ and $K^{*-}\eta'$, which will be labeled
as $-i\Sigma^{K-cVj}$, with $j=1,2,3,4,5,6,7,8$,      {  respectively.    }
The final results read
\begin{eqnarray}
\Sigma^{K-cS1}&& =
\frac{i c_d^2}{2F_0^4} \bigg[\, (3 p^2 + m_K^2- M_\sigma^2)A_0(M_\sigma^2) -( p^2+m_K^2 - M_\sigma^2)A_0(m_K^2)
\nonumber \\&& \qquad \qquad
+(p^2 + m_K^2 - M_\sigma^2)^2 B_0(p^2, M_\sigma^2, m_K^2)   \,\bigg]
\nonumber \\&& \quad
-\frac{i2 c_d c_m}{F_0^4 M_S^2}m_K^2 M_\sigma^2 \bigg[\,
(m_K^2+ p^2 - M_\sigma^2) B_0(p^2, M_\sigma^2, m_K^2)+ A_0(M_\sigma^2) -A_0(m_K^2) \,\bigg]
\nonumber \\&& \quad
+\frac{i2 c_m^2}{F_0^4 M_S^4}m_K^4  \bigg[\,
M_\sigma^4 B_0(p^2, M_\sigma^2, m_K^2)+ (m_K^2+ p^2 + M_\sigma^2)A_0(m_K^2) \,\bigg]\,,
\end{eqnarray}
\begin{eqnarray}
\Sigma^{K-cS2}&& =
\frac{i c_d^2}{F_0^4} \bigg[\, (3 p^2 + m_K^2- M_{\sigma'}^2)A_0(M_{\sigma'}^2) -( p^2+m_K^2 - M_{\sigma'}^2)A_0(m_K^2)
\nonumber \\&& \qquad \qquad
+(p^2 + m_K^2 - M_{\sigma'}^2)^2 B_0(p^2, M_{\sigma'}^2, m_K^2)   \,\bigg]
\nonumber \\&& \quad
-\frac{i4 c_d c_m}{F_0^4 M_S^2}m_K^2 M_{\sigma'}^2 \bigg[\,
(m_K^2+ p^2 - M_{\sigma'}^2) B_0(p^2, M_{\sigma'}^2, m_K^2)+ A_0(M_{\sigma'}^2) -A_0(m_K^2) \,\bigg]
\nonumber \\&& \quad
+\frac{i4 c_m^2}{F_0^4 M_S^4}m_K^4  \bigg[\,
M_{\sigma'}^4 B_0(p^2, M_{\sigma'}^2, m_K^2)+ (m_K^2+ p^2 + M_{\sigma'}^2)A_0(m_K^2) \,\bigg]\,,
\end{eqnarray}
\begin{eqnarray}
\Sigma^{K-cS4}=2\Sigma^{K-cS3}&& =
\frac{i c_d^2}{F_0^4} \bigg[\, (3 p^2 + m_K^2- M_a^2)A_0(M_a^2) -( p^2+m_K^2 - M_a^2)A_0(m_K^2)
\nonumber \\&& \qquad \qquad
+(p^2 + m_K^2 - M_a^2)^2 B_0(p^2,M_a^2,m_K)   \,\bigg]
\nonumber \\&& \quad
-\frac{i4 c_d c_m}{F_0^4 M_S^2}m_K^2 M_a^2 \bigg[\,
(m_K^2+ p^2 - M_a^2) B_0(p^2, M_a^2, m_K^2)+ A_0(M_a^2) -A_0(m_K^2) \,\bigg]
\nonumber \\&& \quad
+\frac{i4 c_m^2}{F_0^4 M_S^4}m_K^4  \bigg[\,
M_a^4 B_0(p^2, M_a^2, m_K^2)+ (m_K^2+ p^2 + M_a^2)A_0(m_K^2) \,\bigg]\,,
\end{eqnarray}
\begin{eqnarray}
\Sigma^{K-cS6}=2\Sigma^{K-cS5}&& =
\frac{i c_d^2}{F_0^4} \bigg[\, (3 p^2 + m_\pi^2- M_\kappa^2)A_0(M_\kappa^2) -( p^2+m_\pi^2 - M_\kappa^2)A_0(m_\pi^2)
\nonumber \\&& \qquad \qquad
+(p^2 + m_\pi^2 - M_\kappa^2)^2 B_0(p^2,M_\kappa^2,m_\pi)   \,\bigg]
\nonumber \\&& \quad
-\frac{i2 c_d c_m}{F_0^4 M_S^2} \bigg\{\,
\big[2m_K^2 M_\kappa^2-(m_K^2-m_\pi^2)(M_\kappa^2-s-m_\pi^2) \big] A_0(M_\kappa^2)
\nonumber \\&&  \qquad \qquad
+\big[-2m_K^2 M_\kappa^2-(m_K^2-m_\pi^2)(m_\pi^2-s-M_\kappa^2) \big] A_0(m_\pi^2)
\nonumber \\&&  \qquad \qquad
+\big[2m_K^2 M_\kappa^2(s+m_\pi^2-M_\kappa^2)
\nonumber \\&&  \qquad \qquad\qquad
-(m_K^2-m_\pi^2)(s^2-M_\kappa^4-m_\pi^4+2m_\pi^2 M_\kappa^2) \big] B_0(p^2, M_\kappa^2, m_\pi^2)\,\bigg\}
\nonumber \\&& \quad
+\frac{i c_m^2}{F_0^4 M_S^4}  \bigg\{\,
\big[4m_K^2(m_K^2-m_\pi^2)M_\kappa^2+(m_K^2-m_\pi^2)^2(m_\pi^2-s-M_\kappa^2) \big] A_0(M_\kappa^2)
\nonumber \\&&
+\big[4m_K^4 (s+m_\pi^2+M_\kappa^2)-4m_K^2(m_K^2-m_\pi^2)(2s+M_\kappa^2)
\nonumber \\&& \quad
+(m_K^2-m_\pi^2)^2(3s+M_\kappa^2-m_\pi^2)  \big] A_0(m_\pi^2)
\nonumber \\&&
+\big[4m_K^4 M_\kappa^4-4m_K^2(m_K^2-m_\pi^2)(s+M_\kappa^2-m_\pi^2)M_\kappa^2
\nonumber \\&&  \quad
+(m_K^2-m_\pi^2)^2(s+M_\kappa^2-m_\pi^2)^2 \big] B_0(p^2, M_\kappa^2, m_\pi^2)
 \,\bigg\},
\end{eqnarray}
\begin{eqnarray}
\Sigma^{K-cS7 }=&& =
\frac{i c_d^2}{2F_0^4} \frac{(c_\theta+2\sqrt2s_\theta)^2}{3} \bigg[\, (3 p^2 + m_\eta^2- M_\kappa^2)A_0(M_\kappa^2) -( p^2+m_\eta^2 - M_\kappa^2)A_0(m_\eta^2)
\nonumber \\&& \qquad \qquad
+(p^2 + m_\eta^2 - M_\kappa^2)^2 B_0(p^2,M_\kappa^2,m_\eta)   \,\bigg]
\nonumber \\&& \quad
-\frac{i c_d c_m}{F_0^4 M_S^2}( c_\theta+2\sqrt2 s_\theta )
\bigg\{
\nonumber \\&&
\big[\, \frac{2( c_\theta+2\sqrt2 s_\theta )}{3}m_K^2 M_\kappa^2
+c_\theta(m_K^2-m_\pi^2)(M_\kappa^2-s-m_\eta^2) \,\big]A_0(M_\kappa^2)
\nonumber \\&&
+ \big[\, -\frac{2( c_\theta+2\sqrt2 s_\theta )}{3}m_K^2 M_\kappa^2
+c_\theta(m_K^2-m_\pi^2)(m_\eta^2-s-M_\kappa^2) \,\big]A_0(M_\eta)
\nonumber \\&&
+ \big[\, \frac{2( c_\theta+2\sqrt2 s_\theta )}{3}m_K^2 (s+m_\eta^2-M_\kappa^2)M_\kappa^2
\nonumber \\&& \qquad
+c_\theta(m_K^2-m_\pi^2)(s^2- m_\eta^4-M_\kappa^4+2m_\eta^2 M_\kappa^2) \,\big] B_0(p^2, M_\kappa^2, m_\eta^2)
 \bigg\}
\nonumber \\&& \quad
+\frac{i c_m^2}{F_0^4 M_S^4}
\bigg\{
\big[\, -2(c_\theta +2\sqrt2 s_\theta)c_\theta m_K^2(m_K^2-m_\pi^2) M_\kappa^2
\nonumber \\&& \qquad
+\frac{3}{2}c_\theta^2(m_K^2-m_\pi^2)^2(m_\eta^2-s-M_\kappa^2) \,\big]A_0(M_\kappa^2)
\nonumber \\&&
+ \bigg[\, \frac{2(c_\theta +2\sqrt2 s_\theta)^2}{3}m_K^4(s+m_\eta^2+M_\kappa^2)
+2(c_\theta +2\sqrt2 s_\theta) c_\theta m_K^2(m_K^2-m_\pi^2)(2s+M_\kappa^2)
\nonumber \\&& \qquad
+\frac{3}{2}c_\theta^2(m_K^2-m_\pi^2)^2(3s+M_\kappa^2-m_\eta^2)  \,\bigg]A_0(M_\eta)
\nonumber \\&&
+ \bigg[\,  \frac{2(c_\theta +2\sqrt2 s_\theta)^2}{3}m_K^4 M_\kappa^4
+2(c_\theta +2\sqrt2 s_\theta) c_\theta m_K^2(m_K^2-m_\pi^2)(s+M_\kappa^2-m_\eta^2)M_\kappa^2
\nonumber \\&& \qquad
+\frac{3}{2}c_\theta^2(m_K^2-m_\pi^2)^2(s+M_\kappa^2-m_\eta^2)^2
 \,\bigg] B_0(p^2, M_\kappa^2, m_\eta^2)
 \bigg\}\,,
\end{eqnarray}
\begin{eqnarray}
\Sigma^{K-cS8 }=&& =
\frac{i c_d^2}{F_0^4} \frac{(2\sqrt2c_\theta-s_\theta)^2}{6}
 \bigg[\, (3 p^2 + m_{\eta'}^2- M_\kappa^2)A_0(M_\kappa^2) -( p^2+m_{\eta'}^2 - M_\kappa^2)A_0(m_{\eta'}^2)
\nonumber \\&& \qquad \qquad
+(p^2 + m_{\eta'}^2 - M_\kappa^2)^2 B_0(p^2,M_\kappa^2,m_{\eta'})   \,\bigg]
\nonumber \\&& \quad
+\frac{i c_d c_m}{F_0^4 M_S^2}( 2\sqrt2 c_\theta-s_\theta)
\bigg\{
\nonumber \\&&
\big[\, \frac{2(s_\theta- 2\sqrt2 c_\theta )}{3}m_K^2M_\kappa^2
+s_\theta(m_K^2-m_\pi^2)(M_\kappa^2-s-m_{\eta'}^2) \,\big]A_0(M_\kappa^2)
\nonumber \\&&
+ \big[\, -\frac{2(s_\theta- 2\sqrt2 c_\theta )}{3}m_K^2 M_\kappa^2
+s_\theta(m_K^2-m_\pi^2)(m_{\eta'}^2-s-M_\kappa^2) \,\big]A_0(M_{\eta'})
\nonumber \\&&
+ \big[\, \frac{2(s_\theta- 2\sqrt2 c_\theta )}{3}m_K^2 (s+m_{\eta'}^2-M_\kappa^2)M_\kappa^2
\nonumber \\&& \qquad
+s_\theta(m_K^2-m_\pi^2)(s^2- m_{\eta'}^4-M_\kappa^4+2m_{\eta'}^2 M_\kappa^2) \,\big] B_0(p^2, M_\kappa^2, m_{\eta'}^2)
 \bigg\}
\nonumber \\&& \quad
+\frac{i c_m^2}{F_0^4 M_S^4}
\bigg\{
\big[\, -2(s_\theta- 2\sqrt2 c_\theta)s_\theta m_K^2 (m_K^2-m_\pi^2) M_\kappa^2
\nonumber \\&& \qquad
+\frac{3}{2}s_\theta^2(m_K^2-m_\pi^2)^2(m_{\eta'}^2-s-M_\kappa^2) \,\big]A_0(M_\kappa^2)
\nonumber \\&&
+ \bigg[\, \frac{2(s_\theta- 2\sqrt2 c_\theta )^2}{3}m_K^4(s+m_{\eta'}^2+M_\kappa^2)
+2(s_\theta- 2\sqrt2 c_\theta)s_\theta m_K^2 (m_K^2-m_\pi^2)(2s+M_\kappa^2)
\nonumber \\&& \qquad
+\frac{3}{2}s_\theta^2(m_K^2-m_\pi^2)^2(3s+M_\kappa^2-m_{\eta'}^2)  \,\bigg]A_0(M_{\eta'})
\nonumber \\&&
+ \bigg[\,  \frac{2(s_\theta- 2\sqrt2 c_\theta )^2}{3}m_K^4 M_\kappa^4
+2(s_\theta- 2\sqrt2 c_\theta)s_\theta m_K^2 (m_K^2-m_\pi^2)(s+M_\kappa^2-m_{\eta'}^2)M_\kappa^2
\nonumber \\&& \qquad
+\frac{3}{2}s_\theta^2(m_K^2-m_\pi^2)^2(s+M_\kappa^2-m_{\eta'}^2)^2
 \,\bigg] B_0(p^2, M_\kappa^2, m_{\eta'}^2)
 \bigg\}\,.
\end{eqnarray}

For the contributions from the vector resonances, the explicit results are
\begin{eqnarray}
\Sigma^{K-cV2}=2\Sigma^{K-cV1}&& =
\frac{i G_V^2}{2F_0^4} \bigg\{\, -( p^2 - m_K^2+ M_\rho^2)A_0(M_\rho^2) -(p^2+m_K^2 - M_\rho^2)A_0(m_K^2)
\nonumber \\&& \qquad \qquad
+\big[( p^2-m_K^2 +M_\rho^2)^2-4p^2 M_\rho^2 \big] B_0(p^2, M_\rho^2, m_K^2)   \,\bigg\}\,,
\end{eqnarray}
\begin{eqnarray}
\Sigma^{K-cV3}&& =
\frac{i G_V^2}{4F_0^4} \bigg\{\, -( p^2 - m_K^2+ M_\omega^2)A_0(M_\omega^2) -(p^2+m_K^2 - M_\omega^2)A_0(m_K^2)
\nonumber \\&& \qquad \qquad
+\big[( p^2-m_K^2 +M_\omega^2)^2-4p^2 M_\omega^2 \big] B_0(p^2, M_\omega^2, m_K^2)   \,\bigg\}\,,
\end{eqnarray}
\begin{eqnarray}
\Sigma^{K-cV4}&& =
\frac{i G_V^2}{2F_0^4} \bigg\{\, -( p^2 - m_K^2+ M_\phi^2)A_0(M_\phi^2) -(p^2+m_K^2 - M_\phi^2)A_0(m_K^2)
\nonumber \\&& \qquad \qquad
+\big[( p^2-m_K^2 +M_\phi^2)^2-4p^2 M_\phi^2 \big] B_0(p^2, M_\phi^2, m_K^2)   \,\bigg\}\,,
\end{eqnarray}
\begin{eqnarray}
\Sigma^{K-cV6}=2\Sigma^{K-cV5}&& =
\frac{i G_V^2}{2F_0^4} \bigg\{\, -( p^2 - m_\pi^2+ M_{K^*}^2)A_0(M_{K^*}^2) -(p^2+m_\pi^2 - M_{K^*}^2)A_0(m_\pi^2)
\nonumber \\&& \qquad \qquad
+\big[( p^2-m_\pi^2 +M_{K^*}^2)^2-4p^2 M_{K^*}^2 \big] B_0(p^2, M_{K^*}^2, m_\pi^2)   \,\bigg\}\,,
\end{eqnarray}
\begin{eqnarray}
\Sigma^{K-cV7 }&& =
\frac{i G_V^2}{4F_0^4} 3c_\theta^2 \bigg\{\, -( p^2 - m_\eta^2+ M_{K^*}^2)A_0(M_{K^*}^2) -(p^2+m_\eta^2 - M_{K^*}^2)A_0(m_\eta^2)
\nonumber \\&& \qquad \qquad
+\big[( p^2-m_\eta^2 +M_{K^*}^2)^2-4p^2 M_{K^*}^2 \big] B_0(p^2, M_{K^*}^2, m_\eta^2)   \,\bigg\}\,,
\nonumber \\&&
\end{eqnarray}
\begin{eqnarray}
\Sigma^{K-cV8 }&& =
\frac{i G_V^2}{2F_0^4} \frac{3s_\theta^2}{2} \bigg\{\, -( p^2 - m_{\eta'}^2+ M_{K^*}^2)A_0(M_{K^*}^2) -(p^2+m_{\eta'}^2 - M_{K^*}^2)A_0(m_{\eta'}^2)
\nonumber \\&& \qquad \qquad
+\big[( p^2-m_{\eta'}^2 +M_{K^*}^2)^2-4p^2 M_{K^*}^2 \big] B_0(p^2, M_{K^*}^2, m_{\eta'}^2)   \,\bigg\}\,.
\end{eqnarray}

\subsection{ The results for $F_\pi^{\rm 1PI}$ in Eq.~\eqref{eq.1PI-Fphi-structure}}
\label{app.1PI-Fpi}

The relevant Feynman diagrams are shown in Fig.~\ref{fig.fphi} and the explicit results for those diagrams will be collected in $T^{\phi}$, with $\phi=\pi, K$.
For the diagram (a), the final expression is
\begin{eqnarray}
T^{\pi-a}=&&\sqrt2\, F_0\, p_\nu\,\big[
 \,\,\, \Frac{\widetilde{F}^2}{F_0^2}
\,\,\, -\,\,\,
 \frac{4\widetilde{L}_{12}}{F_0^2}(p^2-m_\pi^2)+\frac{4\widetilde{L}_{11}}{F_0^2}m_\pi^2
 + \frac{8\widetilde{L}_4}{F_0^2}(2m_K^2+m_\pi^2)
 \nonumber\\ &&
 + \frac{8}{F_0^2}(\widetilde{L}_5+\frac{c_dc_m}{M_S^2})m_\pi^2  \big]\,.
\end{eqnarray}

The result from diagram (b) reads
\begin{eqnarray}
T^{\pi-b} &&= -i\sqrt2\,F_0\,p_\nu\,\bigg[ \frac{4}{3F_0^2}A_0(m_\pi^2) +\frac{2}{3F_0^2}A_0(m_K^2) \bigg]\,,
\end{eqnarray}
which is the same as in $U(3)$ $\chi$PT calculation.

{ The diagram (c)  in Fig~\ref{fig.fphi} receives   }
contributions both from scalar and vector resonances.
Similar to the self-energy case, we take the $\pi^-$ for illustration.
There are five possible combinations of scalar resonance and
pseudoscalar meson running inside the loop, which are exactly the same as in the self-energy calculation:
$\sigma\pi^-$, $\kappa^- K^0$, $\kappa^0 K^-$, $a_0^-\eta$ and
$a_0^-\eta'$, which will be labeled as $T^{\pi-cSj}$, with $j=1,2,3,4,5$,      {  respectively.    }
About the vector, there are four possible combinations: $\rho^-\pi^0$, $\rho^0\pi^-$, $K^{*0} K^-$ and $K^{*-} K^0$ , which
will be labeled as $T^{\pi-cVj}$, with $j=1,2,3,4$,      {  respectively.    }

The final results of these diagrams involving scalar resonances are
\begin{eqnarray}\label{fpis1f}
T^{\pi-cS1}&& =
\frac{-i2\sqrt2\, c_d^2}{F_0^3 p^2}p_\nu \bigg[\, (3 p^2 + m_\pi^2- M_\sigma^2)A_0(M_\sigma^2) -(m_\pi^2+ p^2 - M_\sigma^2)A_0(m_\pi^2)
\nonumber \\&& \qquad \qquad\qquad\quad
+(m_\pi^2+ p^2 - M_\sigma^2)^2 B_0(p^2, M_\sigma^2, m_\pi^2)   \,\bigg]
\nonumber \\&& \quad
+\frac{i4\sqrt2 c_d c_m}{F_0^3 M_S^2 p^2}p_\nu\, m_\pi^2 M_\sigma^2 \bigg[\,
(m_\pi^2+ p^2 - M_\sigma^2) B_0(p^2, M_\sigma^2, m_\pi^2)+ A_0(M_\sigma^2) -A_0(m_\pi^2) \,\bigg]\,, \nonumber \\
\end{eqnarray}
\begin{eqnarray}
&&  T^{\pi-cS2}=T^{\pi-cS3}= \nonumber  \\&& =
\frac{-i \sqrt2\,c_d^2}{F_0^3 p^2}p_\nu \bigg[\, (3 p^2 + m_K^2- M_\kappa^2)A_0(M_\kappa^2) -(m_K^2+ p^2 - M_\kappa^2)A_0(m_K^2)
\nonumber \\&& \qquad \qquad\qquad
+(m_K^2+ p^2 - M_\kappa^2)^2 B_0(p^2, M_\kappa^2, m_K^2)   \,\bigg]
\nonumber \\&& \quad
+\frac{i2\sqrt2 c_d c_m}{F_0^3 M_S^2 p^2}p_\nu  \bigg\{\,
\big[ m_\pi^2 M_\kappa^2+(m_K^2-m_\pi^2)(M_\kappa^2-s-m_K^2) \big] A_0(M_\kappa^2)
\nonumber \\&&  \qquad \qquad
+\big[-m_\pi^2 M_\kappa^2+(m_K^2-m_\pi^2)(m_K^2-s-M_\kappa^2) \big] A_0(m_K^2)
\nonumber \\&&  \qquad \qquad
+\big[ m_\pi^2 M_\kappa^2(s+m_K^2-M_\kappa^2)
+(m_K^2-m_\pi^2)(s^2-M_\kappa^4-m_K^4+2m_K^2 M_\kappa^2) \big] B_0(p^2, M_\kappa^2, m_K^2)\,\bigg\}
\nonumber \\&& \quad
-\frac{i\sqrt2 c_m^2}{ F_0^3 M_S^4 p^2} p_\nu  \bigg\{\,
\big[-2m_\pi^2(m_K^2-m_\pi^2)M_\kappa^2+(m_K^2-m_\pi^2)^2(m_K^2-s-M_\kappa^2) \big] A_0(M_\kappa^2)
\nonumber \\&&
+\big[
2m_\pi^2(m_K^2-m_\pi^2)(2s+M_\kappa^2)
+(m_K^2-m_\pi^2)^2(3s+M_\kappa^2-m_K^2)  \big] A_0(m_K^2)
\nonumber \\&&
+\big[2m_\pi^2(m_K^2-m_\pi^2)(s+M_\kappa^2-m_K^2)M_\kappa^2
+(m_K^2-m_\pi^2)^2(s+M_\kappa^2-m_K^2)^2 \big] B_0(p^2, M_\kappa^2, m_K^2)
 \,\bigg\}\,,
\end{eqnarray}
\begin{eqnarray}
 T^{\pi-cS4} && =
\frac{(c_\theta-\sqrt2 s_\theta)^2}{3}
\times\frac{-i 2\sqrt2\, c_d^2}{F_0^3 p^2} p_\nu \bigg[\, (3 p^2 + m_\eta^2- M_a^2)A_0(M_a^2) -(m_\eta^2+ p^2 - M_a^2)A_0(m_\eta^2)
\nonumber \\&& \qquad \qquad\qquad \qquad
+(m_\eta^2+ p^2 - M_a^2)^2 B_0(p^2, M_a^2, m_\eta^2)   \,\bigg]
\nonumber \\&& \quad
+\frac{i4\sqrt2 c_d c_m}{F_0^3 M_S^2 p^2} p_\nu\,m_\pi^2 M_a^2 \bigg[\,
(m_\eta^2+ p^2 - M_a^2) B_0(p^2, M_a^2, m_\eta^2)+ A_0(M_a^2) -A_0(m_\eta^2) \,\bigg]\,,
\end{eqnarray}
\begin{eqnarray}
 T^{\pi-cS5 } && =
\frac{(\sqrt2 c_\theta+ s_\theta)^2}{3}
\times\frac{-i 2\sqrt2\, c_d^2}{F_0^3 p^2} p_\nu \bigg[\, (3 p^2 + m_{\eta'}^2- M_a^2)A_0(M_a^2) -(m_{\eta'}^2+ p^2 - M_a^2)A_0(m_{\eta'}^2)
\nonumber \\&& \qquad \qquad\qquad \qquad
+(m_{\eta'}^2+ p^2 - M_a^2)^2 B_0(p^2, M_a^2, m_{\eta'}^2)   \,\bigg]
\nonumber \\&& \quad
+\frac{i4\sqrt2 c_d c_m}{F_0^3 M_S^2 p^2} p_\nu\,m_\pi^2 M_a^2 \bigg[\,
(m_{\eta'}^2+ p^2 - M_a^2) B_0(p^2, M_a^2, m_{\eta'}^2)+ A_0(M_a^2) -A_0(m_{\eta'}^2) \,\bigg]\,. \nonumber \\
\end{eqnarray}

For the vector resonances, after an explicit calculation we find that $T^{\pi-cVi}$ is directly related  to
the self-energy function $\Sigma^{Z\pi Vi}$ through
\begin{eqnarray}
T^{\pi-cVi}=-\sqrt2 F_0\,p_\nu\,\frac{1}{p^2}\, \Sigma^{Z\pi Vi}\,, \quad i=1,2,3,4,5.
\end{eqnarray}

\subsection{ The results for $F_K^{\rm 1PI}$ in Eq.~\eqref{eq.1PI-Fphi-structure}}
\label{app.1PI-FK}

It shares the same Feynman diagrams as $F_\pi$ with
different resonances and pseudoscalar mesons running inside the loops in Fig.~\ref{fig.fphi}.
The expression for diagram (a) takes the form
\begin{eqnarray}
T^{K-a}=&&\sqrt2\, F_0\, p_\nu\,\big[
  \,\,\, \Frac{\widetilde{F}^2}{F_0^2}
\,\,\, -\,\,\,
 \frac{4\widetilde{L}_{12}}{F_0^2}(p^2-m_K^2)+\frac{4\widetilde{L}_{11}}{F_0^2}m_K^2
+ \frac{8\widetilde{L}_4}{F_0^2}(2m_K^2+m_\pi^2)
\nonumber \\ &&
+ \frac{8}{F_0^2}(\widetilde{L}_5+\frac{c_dc_m}{M_S^2})m_K^2  \big]\,.
\end{eqnarray}

About the diagram (b), its explicit result is
\begin{eqnarray}
T^{K-b} &&=  -i\sqrt2\,F_0\,p_\nu\,\bigg[ \frac{1}{F_0^2}A_0(m_K^2) + \frac{1}{2F_0^2}A_0(m_\pi^2) +
\frac{c_\theta^2}{2F_0^2}A_0(m_\eta^2)  + \frac{s_\theta^2}{3F_0^2}A_0(m_{\eta'}^2) \bigg] \,.
\end{eqnarray}

For the diagram (c) in Fig.~\ref{fig.fphi}, we take the self-energy for the $K^-$ for illustrating purpose.
Exactly the same as in the self-energy case, there are eight possible combinations of scalar resonance and
pseudoscalar meson running inside the loop: $\sigma K^-$, $\sigma' K^-$, $a_0^0 K^-$, $a_0^-\bar{K}^0$,
$\kappa^-\pi^0$, $\kappa^0\pi^-$, $\kappa^-\eta$ and $\kappa^-\eta'$, which will be labeled as $T^{K-cSj}$,
with $j=1,2,3,4,5,6,7,8$, respectively.
For the vector case, there are also eight possible combinations: $\rho^0 K^-$, $\rho^-\bar{K}^0$, $\omega K^-$, $\phi K^-$,
$K^{*-}\pi^0$, $\bar{K}^{*0}\pi^-$, $K^{*-}\eta$ and $K^{*-}\eta'$, which will be labeled
as $T^{K-cVj}$, with $j=1,2,3,4,5,6,7,8$,      {  respectively.    }
The final expressions for the diagrams involving scalar resonances are
\begin{eqnarray}
T^{K-cS1}&& =
\frac{-i c_d^2}{\sqrt2 F_0^3 p^2} p_\nu \bigg[\, (3 p^2 + m_K^2- M_\sigma^2)A_0(M_\sigma^2) -( p^2+m_K^2 - M_\sigma^2)A_0(m_K^2)
\nonumber \\&& \qquad \qquad\qquad
+(p^2 + m_K^2 - M_\sigma^2)^2 B_0(p^2, M_\sigma^2, m_K^2)   \,\bigg]
\nonumber \\&& \quad
+\frac{i\sqrt2 c_d c_m}{F_0^3 M_S^2 p^2}p_\nu\,m_K^2 M_\sigma^2 \bigg[\,
(m_K^2+ p^2 - M_\sigma^2) B_0(p^2, M_\sigma^2, m_K^2)+ A_0(M_\sigma^2) -A_0(m_K^2) \,\bigg]\,,\nonumber \\
\end{eqnarray}
\begin{eqnarray}
T^{K-cS2}&& =
\frac{-i\sqrt2 c_d^2}{F_0^3 p^2} p_\nu \bigg[\, (3 p^2 + m_K^2- M_{\sigma'}^2)A_0(M_{\sigma'}^2) -( p^2+m_K^2 - M_{\sigma'}^2)A_0(m_K^2)
\nonumber \\&& \qquad \qquad\qquad
+(p^2 + m_K^2 - M_{\sigma'}^2)^2 B_0(p^2, M_{\sigma'}^2, m_K^2)   \,\bigg]
\nonumber \\&& \quad
+\frac{i2\sqrt2 c_d c_m}{F_0^3 M_S^2 p^2}p_\nu\,m_K^2 M_{\sigma'}^2 \bigg[\,
(m_K^2+ p^2 - M_{\sigma'}^2) B_0(p^2, M_{\sigma'}^2, m_K^2)+ A_0(M_{\sigma'}^2) -A_0(m_K^2) \,\bigg]\,, \nonumber \\
\end{eqnarray}
\begin{eqnarray}
T^{K-cS4}=2T^{K-cS3}&& =
\frac{-i\sqrt2 c_d^2}{F_0^3 p^2}p_\nu \bigg[\, (3 p^2 + m_K^2- M_a^2)A_0(M_a^2) -( p^2+m_K^2 - M_a^2)A_0(m_K^2)
\nonumber \\&& \qquad \qquad\qquad
+(p^2 + m_K^2 - M_a^2)^2 B_0(p^2,M_a^2,m_K)   \,\bigg]
\nonumber \\&& \quad \hspace*{-1.5cm}
+\frac{i2\sqrt2 c_d c_m}{F_0^3 M_S^2 p^2}p_\nu\, m_K^2 M_a^2 \bigg[\,
(m_K^2+ p^2 - M_a^2) B_0(p^2, M_a^2, m_K^2)+ A_0(M_a^2) -A_0(m_K^2) \,\bigg]\,, \nonumber \\
\end{eqnarray}
\begin{eqnarray}
&&  T^{K-cS6}=2T^{K-cS5}= \nonumber  \\&& =
\frac{-i \sqrt2\,c_d^2}{F_0^3 p^2}p_\nu \bigg[\, (3 p^2 + m_\pi^2- M_\kappa^2)A_0(M_\kappa^2) -(m_\pi^2+ p^2 - M_\kappa^2)A_0(m_\pi^2)
\nonumber \\&& \qquad \qquad\qquad
+(m_\pi^2+ p^2 - M_\kappa^2)^2 B_0(p^2, M_\kappa^2, m_\pi^2)   \,\bigg]
\nonumber \\&& \quad
+\frac{i2\sqrt2 c_d c_m}{F_0^3 M_S^2 p^2}p_\nu \bigg\{\,
\big[ m_K^2 M_\kappa^2-(m_K^2-m_\pi^2)(M_\kappa^2-s-m_\pi^2) \big] A_0(M_\kappa^2)
\nonumber \\&&  \qquad \qquad
+\big[-m_K^2 M_\kappa^2-(m_K^2-m_\pi^2)(m_\pi^2-s-M_\kappa^2) \big] A_0(m_\pi^2)
\nonumber \\&&  \qquad \qquad
+\big[m_K^2 M_\kappa^2(s+m_\pi^2-M_\kappa^2)
-(m_K^2-m_\pi^2)(s^2-M_\kappa^4-m_\pi^4+2m_\pi^2 M_\kappa^2) \big] B_0(p^2, M_\kappa^2, m_\pi^2)\,\bigg\}
\nonumber \\&& \quad
+\frac{i2\sqrt2 c_m^2}{F_0^3 M_S^4 p^2} p_\nu  \bigg\{\,
\big[-m_K^2(m_K^2-m_\pi^2) M_\kappa^2-\frac{(m_K^2-m_\pi^2)^2}{2}(m_\pi^2-s-M_\kappa^2) \big] A_0(M_\kappa^2)
\nonumber \\&&
+\big[ m_K^2(m_K^2-m_\pi^2)(2s+M_\kappa^2)
-\frac{(m_K^2-m_\pi^2)^2}{2}(3s+M_\kappa^2-m_\pi^2)  \big] A_0(m_\pi^2)
\nonumber \\&&
+\big[m_K^2(m_K^2-m_\pi^2)(s+M_\kappa^2-m_\pi^2)M_\kappa^2
-\frac{(m_K^2-m_\pi^2)^2}{2}(s+M_\kappa^2-m_\pi^2)^2 \big] B_0(p^2, M_\kappa^2, m_\pi^2)
 \,\bigg\}\,,
\end{eqnarray}
\begin{eqnarray}
&&  T^{K-cS7 }= \nonumber  \\&& =
\frac{(c_\theta+2\sqrt2s_\theta)^2}{6}\times\frac{-i \,\sqrt2 c_d^2}{\,F_0^3 p^2}p_\nu
\bigg[\, (3 p^2 + m_\eta^2- M_\kappa^2)A_0(M_\kappa^2) -(m_\eta^2+ p^2 - M_\kappa^2)A_0(m_\eta^2)
\nonumber \\&& \qquad \qquad\qquad
+(m_\eta^2+ p^2 - M_\kappa^2)^2 B_0(p^2, M_\kappa^2, m_\eta^2)   \,\bigg]
\nonumber \\&& \quad
+\frac{i \sqrt2c_d c_m}{\,F_0^3 M_S^2 p^2}p_\nu (c_\theta+2\sqrt2 s_\theta)
\bigg\{
\nonumber \\&&
\big[\, \frac{c_\theta+2\sqrt2 s_\theta}{3}m_K^2 M_\kappa^2
+c_\theta(m_K^2-m_\pi^2)(M_\kappa^2-s-m_\eta^2) \,\big]A_0(M_\kappa^2)
\nonumber \\&&
+ \big[\, -\frac{c_\theta+2\sqrt2 s_\theta}{3}m_K^2 M_\kappa^2
+c_\theta(m_K^2-m_\pi^2)(m_\eta^2-s-M_\kappa^2) \,\big]A_0(M_\eta)
\nonumber \\&&
+ \big[\, \frac{c_\theta+2\sqrt2 s_\theta}{3}m_K^2 (s+m_\eta^2-M_\kappa^2)M_\kappa^2
\nonumber \\&& \qquad
+c_\theta(m_K^2-m_\pi^2)(s^2- m_\eta^4-M_\kappa^4+2m_\eta^2 M_\kappa^2) \,\big] B_0(p^2, M_\kappa^2, m_\eta^2)
 \bigg\}
\nonumber \\&& \quad
-\frac{i \sqrt2c_m^2}{\,F_0^3 M_S^4 p^2} p_\nu  \bigg\{
\nonumber \\&&
\big[\, -c_\theta(c_\theta+2\sqrt2 s_\theta)m_K^2 (m_K^2-m_\pi^2)M_\kappa^2
+\frac{3}{2}c_\theta^2 (m_K^2-m_\pi^2)^2(m_\eta^2-s-M_\kappa^2) \,\big]A_0(M_\kappa^2)
\nonumber \\&&
+ \bigg[\,
c_\theta(c_\theta+2\sqrt2 s_\theta)m_K^2 (m_K^2-m_\pi^2)(2s+M_\kappa^2)
+\frac{3}{2}c_\theta^2 (m_K^2-m_\pi^2)^2(3s+M_\kappa^2-m_\eta^2)  \,\bigg]A_0(M_\eta)
\nonumber \\&&
+ \bigg[\,
c_\theta(c_\theta+2\sqrt2 s_\theta)m_K^2 (m_K^2-m_\pi^2)(s+M_\kappa^2-m_\eta^2)M_\kappa^2
\nonumber \\&& \qquad
+\frac{3}{2}c_\theta^2 (m_K^2-m_\pi^2)^2(s+M_\kappa^2-m_\eta^2)^2
 \,\bigg] B_0(p^2, M_\kappa^2, m_\eta^2)
 \bigg\}\,,
\end{eqnarray}
\begin{eqnarray}
&&  T^{K-cS8 }= \nonumber  \\&& =
\frac{(2\sqrt2c_\theta-s_\theta)^2}{6}\times\frac{-i \,\sqrt2 c_d^2}{\,F_0^3 p^2}p_\nu \bigg[\,
(3 p^2 + m_{\eta'}^2- M_\kappa^2)A_0(M_\kappa^2) -(m_{\eta'}^2+ p^2 - M_\kappa^2)A_0(m_{\eta'}^2)
\nonumber \\&& \qquad \qquad\qquad
+(m_{\eta'}^2+ p^2 - M_\kappa^2)^2 B_0(p^2, M_\kappa^2, m_{\eta'}^2)   \,\bigg]
\nonumber \\&& \quad
-\frac{i \sqrt2c_d c_m}{\,F_0^3 M_S^2 p^2}(2\sqrt2 c_\theta-s_\theta) p_\nu
\bigg\{
\nonumber \\&&
\big[\, \frac{s_\theta-2\sqrt2 c_\theta}{3}m_K^2 M_\kappa^2
+s_\theta(m_K^2-m_\pi^2)(M_\kappa^2-s-m_{\eta'}^2) \,\big]A_0(M_\kappa^2)
\nonumber \\&&
+ \big[\, - \frac{s_\theta-2\sqrt2 c_\theta}{3}m_K^2 M_\kappa^2
+s_\theta(m_K^2-m_\pi^2)(m_{\eta'}^2-s-M_\kappa^2) \,\big]A_0(M_{\eta'})
\nonumber \\&&
+ \big[\,  \frac{s_\theta-2\sqrt2 c_\theta}{3}m_K^2(s+m_{\eta'}^2-M_\kappa^2)M_\kappa^2
\nonumber \\&& \qquad
+s_\theta(m_K^2-m_\pi^2)(s^2- m_{\eta'}^4-M_\kappa^4+2m_{\eta'}^2 M_\kappa^2) \,\big] B_0(p^2, M_\kappa^2, m_{\eta'}^2)
 \bigg\}
\nonumber \\&& \quad
-\frac{i \sqrt2c_m^2}{\,F_0^3 M_S^4 p^2} p_\nu  \bigg\{
\nonumber \\&&\quad
\big[-s_\theta(s_\theta-2\sqrt2c_\theta)m_K^2(m_K^2-m_\pi^2) M_\kappa^2
+\frac{3}{2}s_\theta^2 (m_K^2-m_\pi^2)^2 (m_{\eta'}^2-s-M_\kappa^2) \,\big]A_0(M_\kappa^2)
\nonumber \\&&
+ \bigg[\,
s_\theta(s_\theta-2\sqrt2c_\theta)m_K^2(m_K^2-m_\pi^2)(2s+M_\kappa^2)
+\frac{3}{2}s_\theta^2 (m_K^2-m_\pi^2)^2(3s+M_\kappa^2-m_{\eta'}^2)  \,\bigg]A_0(M_{\eta'})
\nonumber \\&&
+ \bigg[\,
s_\theta(s_\theta-2\sqrt2c_\theta)m_K^2(m_K^2-m_\pi^2)(s+M_\kappa^2-m_{\eta'}^2)M_\kappa^2
\nonumber \\&& \qquad
+\frac{3}{2}s_\theta^2 (m_K^2-m_\pi^2)^2(s+M_\kappa^2-m_{\eta'}^2)^2
 \,\bigg] B_0(p^2, M_\kappa^2, m_{\eta'}^2)
 \bigg\}\,,
\end{eqnarray}

For the vector resonances, we find that $T^{K-cVi}$ is directly related  to
the self-energy function $\Sigma^{K-Vi}$ through
\begin{eqnarray}
T^{K-cVi}=-\sqrt2 F_0\,p_\nu\,\frac{1}{p^2}\, \Sigma^{K-cVi}\,,\quad i=1,2,3,4,5,6,7,8.
\end{eqnarray}

\end{document}